\documentclass[lettersize,journal]{IEEEtran}
\usepackage{amsmath,amsfonts,amssymb}
\usepackage[ruled,lined,linesnumbered]{algorithm2e}
\usepackage{array}
\usepackage{textcomp}
\usepackage{url}
\usepackage{bm}
\usepackage{verbatim}
\usepackage{cite}
\usepackage[colorlinks, citecolor=black]{hyperref}

\usepackage{mathrsfs}
\usepackage{multirow,diagbox}
\usepackage{verbatim}
\usepackage{color}
\usepackage{colortbl}
\usepackage{makecell}
\usepackage{cellspace}
\usepackage{longtable}

\usepackage{amsthm}
\theoremstyle{plain}
\newtheorem{theorem}{Theorem}
\newtheorem{assumption}{Assumption}

\usepackage{booktabs}
\usepackage{array,multirow,multicol,graphicx}
\usepackage{color}
\usepackage{makecell}
\usepackage{siunitx}
\usepackage[x11names,dvipsnames,table]{xcolor} 
\usepackage{colortbl}
\usepackage{multirow}
\usepackage[font=footnotesize]{caption}
\usepackage{subcaption}
\usepackage{lipsum}
\definecolor{mygray}{gray}{0.6}
\definecolor{myblue}{rgb}{0.8,0.85,1}
\usepackage{array}
\newcolumntype{L}[1]{>{\raggedright\let\newline\\\arraybackslash\hspace{0pt}}m{#1}}
\newcolumntype{C}[1]{>{\centering\let\newline\\\arraybackslash\hspace{0pt}}m{#1}}
\newcolumntype{R}[1]{>{\raggedleft\let\newline\\\arraybackslash\hspace{0pt}}m{#1}}

\usepackage{array, tabularx, boldline}
\usepackage{eurosym}
\usepackage{amstext} 
\DeclareRobustCommand{\officialeuro}{%
  \ifmmode\expandafter\text\fi
  {\fontencoding{U}\fontfamily{eurosym}\selectfont e}}

\begin{document}

\title{Privacy-Preserving Federated Unlearning with Certified Client Removal}

\author{Ziyao Liu, Huanyi Ye, Yu Jiang, Jiyuan Shen, Jiale Guo, Ivan Tjuawinata, and Kwok-Yan Lam

%
%
\thanks{Ziyao Liu, Huanyi Ye, Yu Jiang, Jiyuan Shen, Jiale Guo, Ivan Tjuawinata, and Kwok-Yan Lam are with Nanyang Technological University, Singapore.
E-mail: liuziyao@ntu.edu.sg, \{huanyi001, yu012, jiyuan001\}@e.ntu.edu.sg, \{jiale.guo,ivan.tjuawinata\}@ntu.edu.sg, kwokyan.lam@ntu.edu.sg.}
\thanks{Manuscript received November 15, 2023; revised January 16, 2024; accepted April 14, 2024.}}

\markboth{Journal of \LaTeX\ Class Files,~Vol.~14, No.~8, November~2023}%
{Shell \MakeLowercase{\textit{et al.}}: A Sample Article Using IEEEtran.cls for IEEE Journals}

\IEEEpubid{0000--0000/00\$00.00~\copyright~2021 IEEE}

\maketitle
\IEEEpubidadjcol

\begin{abstract}
In recent years, Federated Unlearning (FU) has gained attention for addressing the removal of a client's influence from the global model in Federated Learning (FL) systems, thereby ensuring the ``right to be forgotten" (RTBF). State-of-the-art methods for unlearning use historical data from FL clients, such as gradients or locally trained models. However, studies have revealed significant information leakage in this setting, with the possibility of reconstructing a user's local data from their uploaded information. Addressing this, we propose Starfish, a privacy-preserving federated unlearning scheme using Two-Party Computation (2PC) techniques and shared historical client data between two non-colluding servers. Starfish builds upon existing FU methods to ensure privacy in unlearning processes. To enhance the efficiency of privacy-preserving FU evaluations, we suggest 2PC-friendly alternatives for certain FU algorithm operations. We also implement strategies to reduce costs associated with 2PC operations and lessen cumulative approximation errors. Moreover, we establish a theoretical bound for the difference between the unlearned global model via Starfish and a global model retrained from scratch for certified client removal. Our theoretical and experimental analyses demonstrate that Starfish achieves effective unlearning with reasonable efficiency, maintaining privacy and security in FL systems.

\end{abstract}


\section{Introduction}
\label{sec:introduction}

With the increasing concerns for personal data privacy protection, governments and legislators around the world have enacted significant data privacy regulations, e.g., GDPR \cite{regulation2018general}, to ensure clients ``the right to be forgotten" (RTBF)
These regulations enable individuals to request the removal of their personal data from digital records. In this context, Machine Unlearning (MU) \cite{bourtoule2021machine,cao2015towards,hu2024learn,chen2021machine,hu2024duty,liu2024backdoor,liu2024threats,zhao2024static,qian2023towards} has emerged as a vital enabler of this process, guaranteeing effective and responsible removal of personal data, thereby reinforcing data privacy and ethical data management. Furthermore, Federated Unlearning (FU) has arisen to tackle the challenge of data erasure in the context of federated learning (FL) settings, which has received significant attention recently \cite{cao2023fedrecover,liu2021federaser,halimi2022federated,liu2022right,jiang2024towards,che2023fast,zhang2023fedrecovery,ding2023strategic,tao2024communication}.

The state-of-the-art FU approach, FedRecover \cite{cao2023fedrecover}, eliminates the influence of a target client by leveraging historical information from participating FL clients, including their gradients or locally trained models. More specifically, the server stores all historical information during FL training, including the global model and clients' gradients. Upon receiving the unlearning request from a target client, the server initiates a rollback to the initial global model in the FL training process, which has not been affected by the target client. Based on this initial global model, the server can calibrate the remaining clients' historical gradients in the initial FL round and obtain a new global model. Subsequent calibrations for each FL round are then performed iteratively. This process continues until all historical gradients have been appropriately calibrated, indicating the assumed unlearning of the target client (see Figure \ref{fig:selection} for an illustrative example.). However, as described in \cite{zhu2019deep,liu2022privacy}, an adversarial participant that can access the users' locally trained models, e.g., the server in FedRecover scheme\cite{cao2023fedrecover}, may exploit a Deep Leakage from Gradient (DLG) attack. This type of attack enables the adversary to recover images with pixel-wise accuracy and texts with token-wise matching, resulting in a substantial information leakage of clients' data.

To address such risk, leveraging Two-Party Computation (2PC) techniques \cite{cramer2015secure}, we propose a privacy-preserving federated unlearning scheme Starfish, building upon the state-of-the-art FU approach, to ensure unlearning is conducted in a privacy-preserving manner. More specifically, our proposed scheme ensures privacy preservation by having the clients' historical information secretly shared between two non-colluding servers which jointly simulate the role of the FL server.
When these two servers receive an unlearning request from a target client, they collaboratively execute the FU algorithm using 2PC techniques. 

\IEEEpubidadjcol

Additionally, we have observed that historical information differs in significance to the global model, implying that choosing essential historical data provides enough information for the unlearning process. This suggests that efficient unlearning is possible by focusing on selective historical information rather than information from all historical FL rounds. Building on this concept, we introduce a method for round selection in a privacy-preserving manner, which significantly cuts down the number of unlearning rounds, thus alleviating the substantial costs associated with 2PC operations.

Simultaneously, to enhance the efficiency of the privacy-preserving evaluation of FU, we introduce several 2PC-friendly alternatives for approximating specific operations within the FU algorithm. We then make adaptations and integrate them into the FL training process to further facilitate the 2PC operations during unlearning. Moreover, we also introduce a privacy-preserving protocol to periodically correct errors that accumulate over multiple rounds. Such errors may be due to the estimation errors caused by updating the estimation of the remaining clients in the standard FU process. Alternatively, it may also be caused by approximation errors which occur due to the use of 2PC-friendly alternatives in some of the computations. Furthermore, we establish a theoretical bound, under specific assumptions, to quantify the difference between the unlearned global model obtained through Starfish and the global model obtained through train-from-scratch. This demonstrates that Starfish can achieve federated unlearning with a certified client removal. Theoretical analysis and experimental results show that Starfish can achieve highly effective unlearning in a privacy-preserving manner with reasonable efficiency overheads.

\textbf{Related works.} As discussed in \cite{liu2023survey}, FU approaches aim to unlearn either a target client \cite{halimi2022federated,cao2023fedrecover,liu2021federaser}, or partial data of a target client \cite{li2023federated, che2023fast, xia2023fedme}. Additionally, the target client may actively participate in the unlearning process\cite{zhang2023fedrecovery,wu2023unlearning,gao2022verifi}, or passively engage in the unlearning process \cite{fraboni2022sequential, xia2023fedme, liu2022right}. However, it is worth noting that only a limited number of prior works take privacy-preservation into account. For instance, \cite{zhang2023fedrecovery} concentrates on protecting the privacy of the global model rather than the clients' data. \cite{liu2021revfrf} is primarily focused on the construction of a random forest, \cite{hu2023duty} explores the vulnerabilities associated with unlearning-as-a-service from the perspective of model security, and \cite{liu2020learn} describes a general FU scheme capable of incorporating privacy-enhancing techniques but does not delve into the specifics of optimizing unlearning and privacy preservation. This suggests the need for tailored optimizations which may be utilized in the construction of a privacy-preserving federated unlearning scheme with good trade-offs between privacy guarantees, scheme efficiency, as well as unlearning performance. We note significant differences between Privacy-Preserving Federated Unlearning (PPFU) and Privacy-Preserving Federated Learning (PPFL) \cite{sav2020poseidon,stevens2022efficient,liu2023long,cao2020fltrust,liu2024dynamic,guo2021privacy}. PPFU is designed to protect the privacy of stored historical data across all FL rounds, while PPFL focuses on providing privacy guarantees for each round. Consequently, PPFU aims to achieve a balanced trade-off between storage cost, communication overhead, and model performance. In contrast, PPFL does not need to consider storage costs.


\textbf{Our contributions.} The main contributions of this work are listed as follows.
\begin{enumerate}
    \item We propose a privacy-preserving federated unlearning scheme that enables the server, which is jointly simulated by two non-colluding servers, to unlearn a target client while protecting the data privacy of participating clients.
    \item We describe several 2PC-friendly alternatives to approximate certain operations within the FU algorithm. These alternatives are designed to enhance 2PC efficiency while keeping the unlearning process highly effective.
    \item We enhance efficiency while maintaining unlearning performance by leveraging selective historical information instead of taking information from all historical FL rounds in the unlearning phase, as in the state-of-the-art FU approach. This approach effectively accelerates the unlearning process and mitigates the extensive costs associated with 2PC operations.
    \item
    We establish a theoretical bound that quantifies the difference between the unlearned global model achieved through our proposed scheme and a global model trained from scratch after omitting the requested data from the training data. This bound provides us with a theoretical certified guarantee that the resulting global model is sufficiently close to a model entirely trained-from-scratch with the data held by the target client excluded from the training data.
    \item We conduct a comprehensive experimental analysis of our proposed scheme, evaluating the trade-off between privacy guarantees, scheme efficiency, and unlearning performance.
\end{enumerate}

\textbf{Organisation of the paper.} The rest of the paper is organized as follows. In Section \ref{sec:preliminaries}, we provide the preliminaries with notations used throughout the paper. In Section \ref{sec:overview}, we present the system architecture,  overview of our proposed scheme, a description of the threat model, and the privacy goal. We then proceed to our proposed protocol in Section \ref{sec:proposed_protocols} with detailed theoretical analysis, followed by the experimental evaluation in Section \ref{sec:experimental_valuation}. Finally, we give the conclusions in Section \ref{sec:conclusions}.

\section{Preliminaries and Notations}
\label{sec:preliminaries}

\subsection{Federated Learning \& Unlearning}
\label{sec:fl_fu}
The participants involved in federated learning can be categorized into two categories: (i) a set of $n$ clients denoted as $\mathcal{C}=\{c^1,c^2,\dots,c^n\}$, where each client $c^i \in \mathcal{C}$ possesses its local dataset $\mathcal{D}_i$, and (ii) a central server represented as $S$. A typical FL scheme operates by iteratively performing the following steps until training is stopped \cite{kairouz2021advances}: (a) Local model training: at round $t$, each FL client $c^i$ trains its local model based on a global model $M_t$ using its local dataset $\mathcal{D}_i$ to obtain gradients ${G}_t^i$. (b) Model uploading: each client $c^i$ uploads its gradients ${G}_t^i$ to the central server $S$. (c) Model aggregation: the central server $S$ collects and aggregates clients' models ${G}_t^i$ for $i=1,2,\dots,n$ with some rules, e.g., FedAvg \cite{mcmahan2017communication}, to update the global model ${M_{t+1}}$. (d) Model distribution: the central server $S$ distributes the updated global model ${M_{t+1}}$ to all FL clients.

Building upon the core principles of machine unlearning and the concept of RTBF, federated unlearning aims to enable the global FL model to remove the impact of an FL client or identifiable information associated with the partial data of an FL client, while preserving the integrity of the decentralized learning process \cite{liu2023survey}. 
Federated unlearning can be achieved with different principles involving different participants. In this work, we adhere to the design goal of minimizing client-side computation and communication costs, thus the unlearning step is conducted on the server-side (see Section \ref{sec:design_goals} for more details on the design goals). Therefore, Starfish relies on historical information stored on the server-side for unlearning. The works most closely related to Starfish, with similar design structures, are FedEraser \cite{liu2021federaser} and FedRecover \cite{cao2023fedrecover}. Their performance over plaintext will be compared, and the results are presented in Section \ref{sec:experimental_valuation}.

\subsection{Secure Two-Party Computation}
\label{sec:2pc}


Firstly, we assume that each data is in the form of a real number $x\in \mathbb{R}$ and for the computation, for a predefined finite field $\mathbb{F}_q,$ it has been encoded using a fixed point encoding $\varphi_q:\mathbb{R}\rightarrow \mathbb{F}_q$ \cite{Cat18,MZ17}. A brief discussion on fixed-point encoding and arithmetic can be found in Appendix \ref{app:2pc-fpa}.  

In the following, we will assume that all values have been encoded using the encoding process discussed above to elements of a finite field $\mathbb{F}_q$ for some sufficiently large odd prime $q.$ In constructing our protocols, we assume that the values are secretly shared among $2$ servers, say $\mathcal{S}_1$ and $\mathcal{S}_2.$ Here to secret share a value $v\in \mathbb{F}_q$ among the two parties, we generate a uniformly random value $v_1\in \mathbb{F}_q$ and set $v_2=v-v_1.$ Here $v_1$ and $v_2$ are called the share of $v$ held by $\mathcal{S}_1$ and $\mathcal{S}_2$ respectively. Here we use the notation $[v]\triangleq (v_1,v_2)$ to denote that $v$ is secretly shared using the additive secret sharing scheme as described above. Furthermore, for a vector $\mathbf{v}\in \mathbb{F}_q^n$ and a matrix $M\in \mathbb{F}_q^n,$ we denote by $[\mathbf{v}]$ and $[M],$ the secret shares of $\mathbf{v}$ and $M$ respectively where each entry is secretly shared using the additive secret sharing discussed above. In this work, we assume the existence of various 2PC functionalities, which are discussed in more detail in Appendix \ref{app:2pc-func}. The functionalities that we assume to exist are also summarised in Table \ref{tab:2pcfunc} in Appendix \ref{app:2pc-func}.

In our work, the underlying 2PC scheme that we use is ABY \cite{demmler2015aby}, which, in addition to arithmetic secret sharing discussed above, also uses Boolean secret sharing scheme (where values and shares are binary string) and Yao's Garbled Circuit (where computation is done by having one party generating a garbled version of the required circuit while the other party obliviously evaluate the circuit through the garbled circuit). It can be observed that the required protocols discussed above are proposed in ABY \cite{demmler2015aby,PSSY21}, justifying the use of ABY as our underlying 2PC scheme. We would also like to note that with the use of functionalities proposed in \cite{demmler2015aby,PSSY21}, there are further optimizations on the functionalities discussed in Appendix \ref{app:2pc-funccomp}. The communication round and total communication complexity of such functionalities, as well as several other functionalities utilizing them as building blocks, are summarised in Appendices \ref{app:2pc-funccomp} and \ref{app:2pc-funccomb}.

\section{Design Overview}
\label{sec:overview}

This section presents the system architecture, overview of our proposed protocol, a brief description of the threat model, and our privacy goals.

\subsection{System architecture}

\begin{figure}[htbp!]
    \centering
    \includegraphics[width=0.75\linewidth]{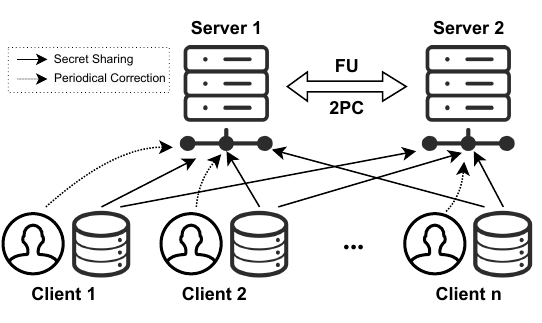}
    \caption{An overview of our system architecture. During the FL training process, all clients share their gradients along with some other assistive information to two non-colluding servers. The server stores all global models. Based on stored historical information, the two servers collaborate in evaluating the FU algorithm using 2PC techniques. Consequently, the two non-colluding servers may choose to recover the final unlearned model for verification of whether the data held by the target client has been successfully unlearned.}
    \label{fig:system_architecture}
\end{figure}

As illustrated in Figure. \ref{fig:system_architecture}, our proposed privacy-preserving FU scheme relies on two non-colluding servers, between which the clients' historical information is secretly shared. When these two servers receive an unlearning request from a target client, they collaboratively execute the FU algorithm using 2PC techniques. Periodically, they correct estimation errors with the remaining clients, all of which lead to the successful unlearning of the data held by the target client\footnote{For simplicity, we assume that the data held by any pair of clients is pairwise disjoint.} in a privacy-preserving manner.

\subsection{Threat model}

Our work considers the following threat model:
We assume the existence of a semi-honest adversary that may corrupt a subset of the participants. Here, among the set of clients and the two servers, the adversary may corrupt a subset
of the clients and at most one server in its effort to acquire information regarding the private data held by the honest parties. Here we note that the adversary may learn all information held by the corrupted parties, however, no participant may deviate from the protocol execution. Importantly, the two servers do not collude with each other. Our consideration is consistent with that of many prior works \cite{xu2020privacy,huang2022cheetah,jayaraman2018distributed,demmler2015aby,he2020secure}.

Note that the vulnerabilities may be exploited by active malicious adversaries who can manipulate the exchanged information. For instance, the server may distribute distinct global models to different clients during specific operations, such as error correction, to distinguish their updates during aggregation. Furthermore, one of the servers may manipulate exchanged information during 2PC to actively obtain more information about the clients' data. Potential countermeasures may involve techniques such as consistency checks \cite{bonawitz2017practical,liu2022efficient} and verifiable computation \cite{damgaard2019new,liu2020mpc}. In this paper, our primary focus is on the semi-honest model, and we defer the investigation of the active malicious threat model to future work.

\subsection{Design goals}
\label{sec:design_goals}
\textbf{Privacy preservation.} Our work aims to protect the privacy of clients' gradients during the whole FU process, specifically by ensuring that each client's gradients remain private to any other participants. This is in place to prevent adversaries from exploiting DLG attacks \cite{zhu2019deep} as discussed earlier. We would like to note that we aim to protect the users' gradient after each local update, the global models are publicly accessible to all participants and are transmitted over plaintext. This design facilitates the efficient execution of certain 2PC operations.

\textbf{Efficient unlearning.} The unlearning process through Starfish should be efficient within a 2PC setting. Considerations for optimizations should be given to accelerate the unlearning process and mitigate the extensive costs associated with 2PC operations. Our objective is to formulate a cost-effective privacy-preserving FU method, ensuring the server can unlearn a target client within a reasonable cost. Additionally, the proposed FU methods should incur minimal client-side computation and communication costs.

\textbf{Certified client removal.} The global FL model unlearned through Starfish should closely match the performance of the train-from-scratch baseline. Furthermore, the difference between the unlearned global model obtained through Starfish and the global model obtained through train-from-scratch should be bounded under certain assumptions.

\section{Proposed Protocols}
\label{sec:proposed_protocols}

Note that Starfish achieves privacy preservation through secret sharing and 2PC techniques. Additionally, Starfish improves efficiency while preserving unlearning performance by employing (i) selective utilization of historical information instead of the consideration of information from all historical FL rounds, and (ii) the utilization of 2PC-friendly alternatives in approximating specific operations within the FU algorithm. In this section, we present detailed designs for these strategies as part of the comprehensive description of the Starfish scheme. Furthermore, we will provide a theoretical analysis of the difference between the global model obtained through Starfish and the one obtained through train-from-scratch strategy. Note that all the algorithms described can be employed to unlearn a set of target clients. For the sake of simplicity and the readability of the discussion, we present the description of the unlearning step where there is only one target client.

\subsection{Selecting historical rounds}

In this section, we will describe how the servers select historical rounds for FU and how to do it in a privacy-preserving manner.

\subsubsection{Round selection}\label{sec:RounSel}
As described earlier, we have observed that the contribution of the target client varies in importance to the global model, suggesting that selecting essential historical information offers sufficient information for the unlearning process. This indicates that highly effective unlearning can still be achieved based on selective historical information rather than information from all historical FL rounds. Upon receiving the unlearning request from a target client, the server assesses the contribution of this target client to each historical round by calculating the cosine similarity between its local update, i.e., gradients, and the global update, i.e., the aggregated gradients from all participating clients or the difference between the current global model and the one from the previous FL round, in each round. Note that cosine similarity is a widely used metric to measure the angle between two vectors, providing a measure of the updating direction similarity between a local model and the global model \cite{cao2021fltrust,zhang2023fltracer}.

\begin{figure}[htbp!]
    \centering
    \includegraphics[width=0.9\linewidth]{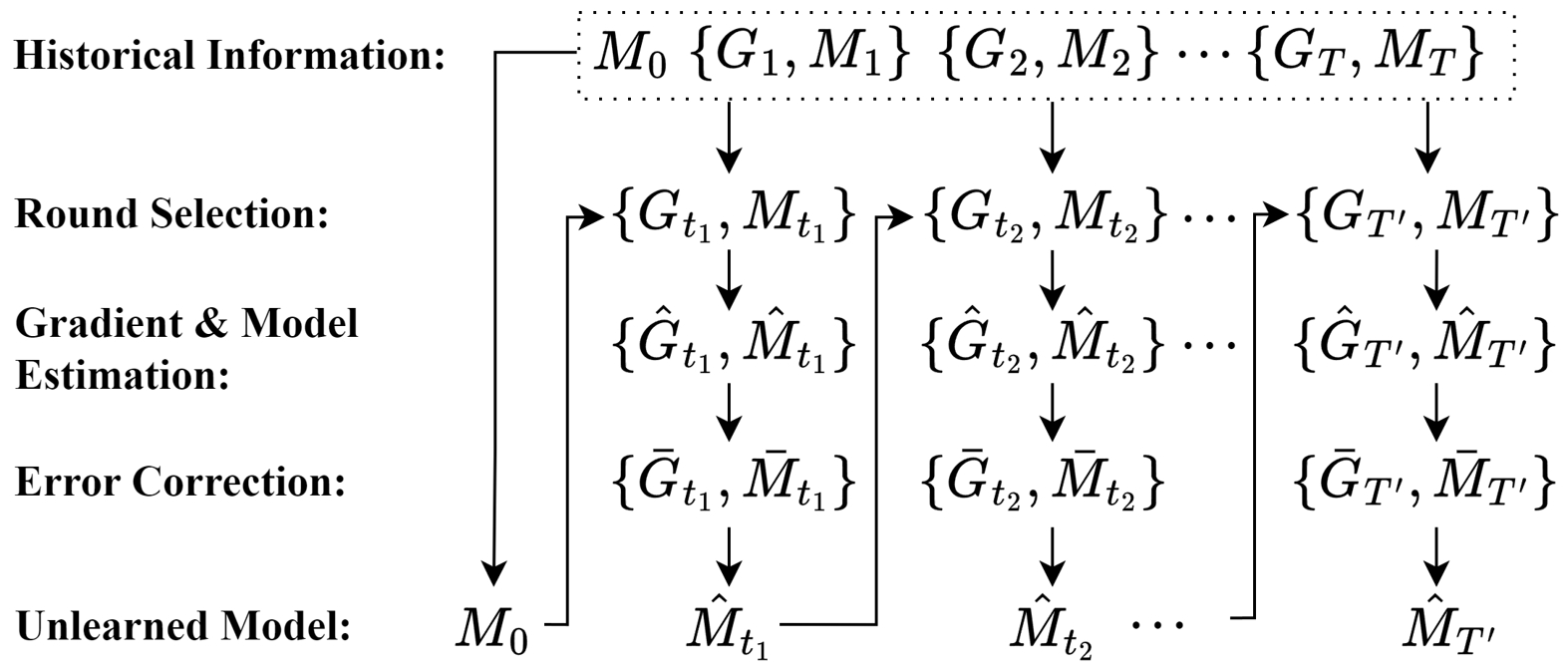}
    \caption{An illustration of the Starfish scheme. The servers store the initial model, historical gradients from all clients, and global models. Upon receiving the unlearning request from a target client, the server selects some historical rounds based on the historical gradients of the target client. Based on the selected rounds, the server obtains historical gradients excluding those from the target client, along with the corresponding global models. Then the server calibrates those selected gradients and global models with estimation and error correction in an iterative style.}
    \label{fig:selection}
\end{figure}

More specifically, assume that the system receives the unlearning request at round $T$ from a target client $c^t$. At this point, the server has stored all clients' historical gradients, denoted by $\{G_i\}^T=\{G_i|i=1,\dots,T\}$, where each $G_i$ consists of the gradients from all $n$ clients at round $i$, i.e., $G_i=\{g_i^j\}^n=\{g_i^j|j=1,\dots,n\}$, along with the global models, denoted by $\{M_i\}^T=\{M_i|i=1,\dots,T\}$ for $T$ rounds and the initial global model $M_0$.
As has been previously discussed, in the actual scheme, the local gradients of the users $g_i^j$ are stored in a secret sharing form $[g_i^j]$ while the global model $M_i$ is stored in the clear. We note that in the initial description of the protocol, for simplicity of the argument, we will omit such distinction and they will be described differently during the protocol specification. Then, the server can assess the contribution of the target client $c^t$ to each historical round $i$ by computing the cosine similarity $\tau_i^t$ between its local update $g_i^t$ and the global update $M_i - M_{i-1}$, as follows:

\begin{equation}
\label{equ:cosine_similarity}
    \tau_i^t = \frac{\left\langle g_i^t, M_i - M_{i-1} \right\rangle}{\Vert g_i^t \Vert \cdot \Vert M_i - M_{i-1} \Vert} 
\end{equation}

where $\left\langle,\right\rangle$ is the scalar product operator. After that, the server can summarize the cosine similarities of the target client $\{\tau_i^t\}^T=\{\tau_i^t|i=1,\dots,T\}$ and sort them to select $T^{\prime}=\lceil \sigma T \rceil$ rounds with the largest cosine similarity $\tau_i^t$, where $\sigma$ represents a hyperparameter termed as the selection rate.  As a result, the server obtains $T^{\prime}$ selected historical gradients $\{G_i\}^{T^{\prime}}$ excluding those from the target client, where $G_i=\{g_i^j\}^{n-1}$, along with selected historical global models $\{M_i\}^{T^{\prime}}$, which are illustrated as $\{G_{t_1},\dots,G_{T^{\prime}}\}$ and $\{M_{t_1},\dots,M_{T^{\prime}}\}$, respectively, in Figure. \ref{fig:selection}. A detailed description is presented in Algorithm \ref{alg:round_selection}. We note that Algorithm \ref{alg:round_selection} is constructed while taking the optimizations discussed in the following sections into account.

\subsubsection{Optimizations on 2PC}
Since in the Starfish scheme, the gradients of the target client $g_i^t$ are secretly shared among two servers, computing the cosine similarities $\tau_i^t$ as given in Equation \ref{equ:cosine_similarity} is conducted using 2PC techniques. Specifically, since global models are public to all FL participants, including both clients and the server, $M_i - M_{i-1}$ is over plaintext. Therefore, operations of two-party secure computation over two ciphertexts, i.e., secret shares, in evaluating Equation \ref{equ:cosine_similarity} involve only secure square root for computing the 
$\ell_2$-norm $||g_i^t||$ and secure division for computing $\frac{\left\langle g_i^t, M_i - M_{i-1} \right\rangle}{\Vert g_i^t \Vert \cdot \Vert M_i - M_{i-1} \Vert}$. After that, a secure sorting operation on $\{\tau_i^t\}^T$ is necessary to select $T^{\prime}$ rounds from $\{G_i\}^T$, resulting in $\{G_i\}^{T^{\prime}}$.

\begin{algorithm}[h!]
\SetAlgoNoEnd
\caption{Secure Round Selection (SecRS)}
\label{alg:round_selection}
\KwIn{Historical gradients $\{[G_i]\}^T$ with their $\ell_2$-norm values $[\{[||G_i||]\}^T]$, historical global models $\{M_i\}^T$, the initial global model $M_0$, the number of historical rounds $T$, the number of selected rounds $T^{\prime}$, the switching threshold $T_{\delta}$, the target client $c^t$.}
\KwOut{Selected historical gradients $\{[G_i]\}^{T^{\prime}}$ along with corresponding selected historical global models $\{M_i\}^{T^{\prime}}$.}
Obtain $\{[g_i^t]\}^T,\{||[g_i^t]||\}^T$ from $\{[G_i]\}^T,\{||[G_i]||\}^T$\;
$[\mathcal{L}]\leftarrow \epsilon$\; \tcp{Initialize $\mathcal{L}$ as an empty list}

\eIf{$T \leq T_{\delta}$}{
\For{$i \leftarrow 1$ \KwTo $T$ \tcp{\textit{Method 1}}}{
$[u_i]\leftarrow \text{SecSP}([g_i^t],M_i-M_{i-1})$ \;
$[v_i]\leftarrow \text{SecMul}([\|g_i^t\|],\|M_i-M_{i-1}\|)$ \;
$[\mathcal{L}]\leftarrow [\mathcal{L}]\|(i,[u_i],[v_i])$ \tcp{Append $(i,[u_i],[v_i])$ to $[\mathcal{L}]$}
}
$[\mathcal{L}']\leftarrow \text{SecSrt}([\mathcal{L}],\text{SecGE2})$\; \tcp{Sort $\mathcal{L}$ based on the ``$\succcurlyeq$'' rule defined in $\text{SecGE2}$ and store it in $[\mathcal{L}']$}
}
{
\For{$i \leftarrow 1$ \KwTo $T$ \tcp{\textit{Method 2}}}{
$[u_i] \leftarrow \text{SecSP}([g_i^t],M_i - M_{i-1})$\;
$[v_i] \leftarrow \text{SecMul}([||g_i^t||],||M_i - M_{i-1}||)$\;
$[\tau_i^t] \leftarrow \text{SecDiv}([u_i],[v_i])$\;
$[\mathcal{L}]\leftarrow [\mathcal{L}]\|(i,[\tau_i^t])$
}
$[\mathcal{L}'] \leftarrow \text{SecSrt}(\{[\tau_i^t]\}^T,\text{SecGE})$\;
}
$\mathbf{d}\leftarrow \epsilon$\;
\For{$i\leftarrow 1$ \KwTo $T'$}
{$\mathbf{d}\leftarrow \mathbf{d}\|\text{SecRec}([\mathcal{L}'[i,0]])$\; \tcp{The index of the $i$-th largest value is recovered from the first element of the $i$-th entry of the sorted $\mathcal{L}'$}}
Obtain $\{[G_i]\}^{T^{\prime}}\triangleq \{[G_{d[i]}]\}_{i=1,\cdots, T'}, \{M_i\}^{T^{\prime}}\triangleq \{[M_{d[i]}]\}_{i=1,\cdots, T'}$ from $\{[G_i]\}^T$ based on $\mathbf{d}$\;
\KwRet{$\{[G_i]\}^{T^{\prime}}, \{M_i\}^{T^{\prime}}$}
\end{algorithm}

\textbf{Replacing secure square root with pre-computation.} It is essential to note that, as $g_i^t$ is uploaded to the server before the updated global model $M_i$ is distributed to all clients, $\tau_i^t$ cannot be calculated by the target client $c^t$ before the global model is updated. Indeed, we can require the target client to calculate $\tau_i^t$ after receiving the updated global model $M_i$, and this value can be secretly shared between two servers to avoid evaluating $\tau_i^t$ using 2PC, thus improving efficiency during round selection. However, if the target client is an actively malicious participant, as described in FedRecover \cite{cao2023fedrecover}, it may manipulate its gradient $g_i^t$ to execute poisoning attacks by injecting backdoors. Allowing the target client to locally compute $\tau_i^t$ might enable it to degrade the performance of unlearning, potentially leading to the persistence of backdoors that cannot be effectively removed through the unlearning process. Therefore, in the Starfish scheme, secure computation on $\tau_i^t$ is conducted on the server-side to make it suitable for various application scenarios. Nevertheless, for improved efficiency, each client $c^j$ can be directed to calculate $||g_i^j||$ locally before receiving the updated global model $M_i$, eliminating the need for a secure square root during round selection. In each FL round $i$, each client $c^j$ is then required to share its gradients $g_i^j$ along with the $\ell_2$-norm $||g_i^j||$ between the two servers.

\textbf{Replacing secure division with secure multiplication.}
As previously outlined, the process of round selection involves finding the $T'$ largest pairs $([u_i],[v_i])$ where $u_i\triangleq \langle [g_i^t],M_i-M_{i-1}\rangle, v_i\triangleq \text{SecMul}([\|g_i^t\|],\|M_i-M_{i-1}\|),$ and we say $(u_i,v_i)$ is larger than $(u_j,v_j),$ denoted by $(u_i,v_i)\succcurlyeq(u_j,v_j)$ if $\frac{u_i}{v_i}\geq \frac{u_j}{v_j} \Leftrightarrow u_iv_j \geq v_iu_j.$ Previously, this calculation required calls to the secure division functionality to calculate $\frac{u_i}{v_i}$ which can then be compared. In this section, we discuss a possible optimization technique that performs the same functionality without the need to call the secure division functionality which, in general, has a larger complexity. Our optimization, which we call \textit{Method 1} is done as follows. Given two pairs $(u_i,v_i)$ and $(u_j,v_j),$ we utilize $\text{SecGE2}([(u_i,v_i)],[(u_j,v_j)])$ which yields $1$ if $(u_i,v_i)\succcurlyeq(u_j,v_j)$ and $0$ otherwise. This can be realized by first calling $[w_i]=\text{SecMul}(u_i,v_j), [w_j]=\text{SecMul}(u_j,v_i)$ and returns the output of $\text{SecGE}([w_i],[w_j]).$ In order to facilitate comparison, we call the original approach \textit{Method 2:} identify the $T'$ largest pairs through the initial computation of $\tau_i^t=\frac{u_i}{v_i}$ for each $i=1,\cdots, T$, followed by sorting the list based on $\tau_i^t.$ Note that in this case, for a list of $T$ pairs where we want to return the $T'$ largest pairs, assuming that $n_{comp}(T)$ comparisons are done to achieve such goal, which is a value that depends on the value of $T,$ comparing the first method and the second, the first method requires additional $2n_{comp}(T)$ calls of $\text{SecMul}$ while the second method requires additional $T$ calls of $\text{SecDiv}.$  Hence, we can identify a switching threshold $T_{\delta}$ such that when $T<T_{\delta},$ the first method is more efficient than the second one. More specifically, let $T_{\text{SecMul}}$ and $T_{\text{SecDiv}}$ be the complexity required in a call of the $\text{SecMul}$ and $\text{SecDiv}$ protocols respectively. Then the first method should be used if $T\cdot T_{\text{SecDiv}}\geq 2n_{comp}(T)\cdot T_{\text{SecMul}}.$ Note that if they require the same number of communication rounds, the first approach is preferred since secure division introduces a larger expected precision error. The selection of $T_{\delta}$ is discussed in Section \ref{sec:theoretical_analysis}, while detailed descriptions of secure round selection can be found in Algorithm \ref{alg:round_selection}.

Note that all operations in Algorithm \ref{alg:round_selection} are conducted between two servers and do not require interaction with clients. Since round indexes are public to both clients and the server, steps 1 and 21 can be straightforwardly executed.

\subsection{Estimating updates}
In this section, we will describe how the servers estimate gradients and global models for unlearning and how to do it in a privacy-preserving manner.

\subsubsection{Update estimation}

In the state-of-the-art solution, FedRecover \cite{cao2023fedrecover}, the server estimates model updates from remaining clients, eliminating the need for local client computation and transferring the computation task to the server. This adaptability makes it well-suited for cross-device FL settings, especially when clients are resource-constrained mobile devices. In Starfish, we follow a similar design goal of minimizing client-side computation and communication costs, thus conducting the estimation of model updates on the two-servers-side. Specifically, after the servers have selected the rounds from which the historical information are obtained, i.e., $\{G_i\}^{T^{\prime}}$ and $\{M_i\}^{T^{\prime}}$, they can calculate the estimated gradients $\{\hat{G}_i\}^{T^{\prime}}$ and the global models $\{\hat{M}_i\}^{T^{\prime}}$. Adapted from FedRecover \cite{cao2023fedrecover}, the servers calculate the estimated gradients and global model as follows, derived from the L-BFGS algorithm \cite{byrd1994representations}:

\begin{equation}
\label{equ:model_estimation}
\begin{aligned}
    \hat{M}_{t_i+1} &= \hat{M}_{t_{i}} -\eta \hat{G}_{t_{i}} \\
    &= \hat{M}_{t_{i}} -\eta_u \mathcal{H}_{t_{i}}^{-1} G_{t_{i}}
\end{aligned}
\end{equation}

where $\eta_u$ is the unlearning rate. The inverse Hessian matrix $\mathcal{H}_{t_{i}}^{-1}$ in Equation \ref{equ:model_estimation} is recursively approximated as follows:

\begin{equation}
\label{equ:hessian_approximation}
    \mathcal{H}_{t_{i}+1}^{-1} = V_{t_{i}}^{tr} \mathcal{H}_{t_{i}}^{-1} V_{t_{i}} + \rho_{t_{i}} \Delta G_{t_{i}} \Delta G_{t_{i}}^{tr}
\end{equation}

where $tr$ represents the matrix transpose operation, and
\begin{equation}
\label{equ:hessian_iteration}
\begin{aligned}
    & \Delta G_{t_{i}} = \hat{G}_{t_i}-{G}_{t_{i}}, \\
    & \Delta M_{t_{i}} = \hat{M}_{t_i}-{M}_{t_{i}}, \\
    & \rho_{t_{i}} = (\Delta G_{t_{i}}^{tr}\Delta M_{t_{i}})^{-1}, \mathrm{~and~} \\
    &  V_{t_{i}} = I-\rho_{t_{i}} \Delta G_{t_{i}} \Delta M_{t_{i}}^{tr}
\end{aligned}
\end{equation}

Here, $M_{t_0}$ is set to be the initial global model $M_0$, while $\hat{M}_{t_i}$ and $\hat{G}_{t_i}$ are calculated by the remaining clients, following an aggregation rule $\mathcal{A}$, e.g., FedAvg \cite{mcmahan2017communication}. Here we denote by $\text{SecAdd}_{\mathcal{A}}$ the secure realization of the aggregation based on $\mathcal{A},$ which may involve calls of $\text{SecAdd}$ and $\text{SecMul}$.
These values are then secretly shared between the two servers to preserve the privacy of various private data. A detailed description is presented in Algorithm \ref{alg:update_estimation}.

\begin{algorithm}[h!]
\SetAlgoNoEnd
\caption{Secure Update Estimation (SecUE)}
\label{alg:update_estimation}
\KwIn{Selected historical gradients $\{[G_{i}]\}^{T^{\prime}}$, Selected historical global models $\{M_i\}^{T^{\prime}}$, the initial model $M_0$, the unlearning rate $\eta_u$, the buffer size $B$, the target client $c^t$, the total number of unlearning rounds $T^{\prime}$.}
\KwOut{Estimated historical gradients $\{[\hat{G}_i]\}^{T^{\prime}}$ along with corresponding global models $\{\hat{M}_i\}^{T^{\prime}}$.}
Initialize $\left(\mathcal{H}_0^j\right)^{-1}$ to a random positive definite matrix of the required dimension\;
The clients compute a $B$ set of $\Delta M_{t_i}$, and $\Delta G_{t_i}$ which are then secretly shared between servers\;
\For{$i \leftarrow T^{\prime}-B$ \KwTo $T^{\prime}$}{
$[\hat{G}_{t_i}^{agg}] = [\mathbf{0}]$\;
\For{$j \leftarrow 1$ \KwTo $n$, $j \neq t$ in parallel}{
$[\Delta G_{t_{i}}^j] = \text{SecAdd}([\hat{G}_{t_i}^j],-[{G}_{t_{i}}^j])$\;
$\Delta M_{t_{i}}^j = \hat{M}_{t_i}^j-{M}_{t_{i}}^j$\;
$[\rho_{t_{i}}^j] = \text{SecMI}([\Delta G_{t_{i}}^j]^{tr}\Delta M_{t_{i}}^j)$\;
$[V_{t_{i}}^j] = \text{SecAdd}(I,-\text{SecMul3}([\rho_{t_{i}}^j],$ $[\Delta G_{t_{i}}^j],$ $\left({\Delta M_{t_i}^j}\right)^{tr}))$\;
$[{\mathcal{H}_{{t_i}+1}^j}^{-1}] = \text{SecAdd}(\text{SecMul3}([V_{t_{i}}^j]^{tr},$
$[{\mathcal{H}_{t_{i}}^j}^{-1}],[V_{t_{i}}^j]),\text{SecMul3}([\rho_{t_{i}}^j],[\Delta G_{t_{i}}^j],$ $[\Delta G_{t_{i}}^j]^{tr}))$\;
$[\hat{G}_{t_i+1}^j] = \text{SecMul}([{\mathcal{H}_{t_{i}+1}^j}^{-1}],[{G}_{t_{i}}])$\;
}
$[\hat{G}_{t_i}^{agg}]=\text{SecAdd}_{\mathcal{A}}([\hat{G}_{t_i+1}])$\;
Recover $\hat{G}_{t_i}^{agg} \leftarrow \text{SecRec}([\hat{G}_{t_i}^{agg}])$\;
}
$\hat{M}_{t_i+1} = \hat{M}_{t_{i}} -\eta_u \hat{G}_{t_{i}}$\;
\KwRet{$\{[\hat{G}_i]\}^{T^{\prime}}\triangleq \{[\hat{G}_{t_i}]:i=1,\cdots, T'\},\{\hat{M}_i\}^{T^{\prime}}\triangleq \{[\hat{M}_{t_i}]:i=1,\cdots, T'\}$}
\end{algorithm}

\subsubsection{Optimizations on 2PC}
Now we proceed to explain how to evaluate the update estimation through the equations provided above, utilizing 2PC techniques.

\textbf{Evaluating matrix inversion.} We observe that in Equation \ref{equ:model_estimation},\ref{equ:hessian_approximation}, and \ref{equ:hessian_iteration}, only the evaluation of $\rho_{t_{i}}$ involves calculating the matrix inversion, while the remaining computations can be obtained through SecAdd and SecMul protocols. Therefore, we propose a 2PC protocol to securely compute the inverse of a matrix, as given in Algorithm \ref{alg:matrix_inversion} in Appendix \ref{app:SecMI}. A theoretical analysis of this additional cost will be presented in Section \ref{sec:theoretical_analysis}, complemented by experimental evaluation in Section \ref{sec:experimental_valuation}.

\textbf{Reducing memory consumption through approximation with a trade-off.}
Additionally, to avoid the storage of $\mathcal{H}_{t_{i}}^{-1}$, which consumes a large amount of memory for large-scale FL models, the server stores a recent set $B$ of $\Delta G=\{[\Delta G_{t_{i}}]\}^B$ and $\Delta M=\{[\Delta M_{t_{i}}]\}^B$ in a buffer, rather than those from all rounds, which are then used to approximately evaluate Equation \ref{equ:model_estimation}. This strategy is derived from \cite{byrd1994representations}, to which interested readers may refer for more details. However, while this strategy reduces memory consumption, it introduces a trade-off by increasing communication costs due to 2PC operations for iteratively calculating $\mathcal{H}_{t_{i}}^{-1}$ in every unlearning round between two servers. Theoretical and empirical analyses of this trade-off concerning the selection of the buffer size $B$ are provided in Section \ref{sec:theoretical_analysis} and Section \ref{sec:experimental_valuation}, respectively.

\subsection{Correcting errors}
\label{sec:correcting_errors}
In this section, we will describe how the servers and remaining clients jointly correct the cumulative errors caused by the approximation of the L-BFGS algorithm and 2PC alternatives, and how to do it in a privacy-preserving manner.

\subsubsection{Error correction}
Due to the approximation operations involved in update estimation and round selection within the Starfish scheme, errors may accumulate across multiple rounds during the unlearning process. This could potentially lead to a less accurate unlearned global model. Therefore, we propose an error correction method to mitigate this challenge. In FedRecover \cite{cao2023fedrecover}, error correction is implemented through ``abnormality fixing'', where the server instructs the remaining clients to compute their exact model update when at least one coordinate of the estimated global model update surpasses a predefined threshold. This threshold is determined by the fact that a fixed fraction of all historical gradients, excluding those from target clients, surpasses this threshold. The error correction method in Starfish differs from the one employed in FedRecover \cite{cao2023fedrecover}, specifically designed to be 2PC-friendly.

More specifically, different thresholds are set for different clients in Starfish. For a remaining client $c^r$, a threshold $\delta_i^r$ is computed for the $i$-th round such that $\alpha$ fraction, termed as the tolerance rate, of its gradients, are greater than $\delta_i^r$ in that round. The threshold for client $c^r$ is then determined as $\delta^r=\max(\{\delta_i^r\}^T).$ During each unlearning round in Starfish, at intervals of every $T_c$ unlearning rounds, the servers instruct each remaining client $c^r$ to compute its exact gradients $\{[\bar{G}_{t_i}]\}.$ For each of such remaining client, if at least one coordinate of its gradients is larger than its threshold $\delta^r$, the correction flag for the client $c^r$  is set to be $1$. We note here that in order to avoid the need of secure square root computation, instead of comparing the norm for each coordinate of $[\bar{G}_{t_i}]$ with $\delta^j,$ noting that both values are non-negative, we may compare their squared values, as described in Step $4.$ These corrected gradients $\{[\bar{G}_{t_i}]\}$ are then secretly shared between the two servers to replace the corresponding estimated gradients. A detailed description of securely checking the threshold is presented in Algorithm \ref{alg:threshold_checking}, and the full description of the error correction process can be found in Algorithm \ref{alg:starfish}.

\begin{algorithm}
\SetAlgoNoEnd
\caption{Secure Threshold Checking (SecTC)}
\label{alg:threshold_checking}
\KwIn{Estimated gradients $[\hat{G}_{t_i}]$ with size $m$, a set of threshold $\{[\delta^j]\}^{n-1}$, remaining clients $\{c^j\}^{n-1}$, the target client $c^t$.}
\KwOut{The correction flags $f_{t_i}$.}
\For{$j \leftarrow 1$ \KwTo $n$, $j \neq t$ in parallel}{
$[e_{t_i}^j]=[\mathbf{0}],[f_{t_i}^j]=[\mathbf{0}]$\;
\For{$k \leftarrow 1$ \KwTo $m$ in parallel}{
$[e_{t_i}^j(k)]=\text{SecGE}(\text{SecSP}([\hat{G}_{t_i}^j(k)],[\hat{G}_{t_i}^j(k)]),$ $\text{SecMult}([\delta^j],[\delta^j]))$\;
$[e_{t_i}^j]=\text{SecAdd}([e_{t_i}^j],[e_{t_i}^j(k)])$\;
}
$[f_{t_i}^j]=\text{SecGE}([e_{t_i}^j],1)$
}
Recover $\{f_{t_i}^j\}^{n-1} \leftarrow \text{SecRec}(\{[f_{t_i}^j]\}^{n-1})$\;
\KwRet{$f_{t_i}$}
\end{algorithm}

\subsubsection{Optimizations on 2PC} As mentioned earlier, in FedRecover \cite{cao2023fedrecover}, finding a threshold involves determining a fixed fraction of all historical gradients, excluding those from target clients, that surpasses the threshold. However, given that the gradients are secretly shared between two servers, which requires 2PC operations, implementing this in Starfish is impractical. Therefore, we propose an optimized alternative for threshold checking.

\textbf{Minimizing the number of secure comparisons.} Attempting a straightforward conversion of the threshold determination method in FedRecover \cite{cao2023fedrecover} to a privacy-preserving version is inefficient. This is attributed to the substantial number of secure comparison operations required for sorting. Therefore, Starfish adopts a strategy of setting different thresholds for individual clients. By computing a threshold involving secure comparison operations over gradients from only one client, whose dimension is $n$ times smaller than what is considered in \cite{cao2023fedrecover}, the number of secure comparison operations is significantly reduced. This results in a noteworthy improvement in efficiency, particularly in communication-intensive 2PC. It is important to highlight that this approach introduces an extra 2PC operation for a secure maximum to determine the largest threshold for each client from the thresholds for all historical rounds, as explained earlier. An analysis of the complexity comparison is presented in Section \ref{sec:theoretical_analysis}, accompanied by corresponding empirical results in Section \ref{sec:experimental_valuation}.

\subsection{Putting it all together}
Building upon the algorithms and strategies outlined earlier, we can present the Starfish scheme for privacy-preserving federated learning. Protocols in the Starfish scheme operate between two non-colluding servers, $S_1$ and $S_2$, and a set of $n-1$ remaining clients to unlearn the target client $c^t$. We divide the Starfish scheme into two stages. \textbf{Stage I}: throughout the FL process, in each round, all clients are directed to upload the secret sharing of their gradients along with assistive information to the two servers. Using 2PC techniques, the servers aggregate the clients' gradients to derive the updated global model. This updated global model is then distributed to all clients for training in the subsequent federated learning round. \textbf{Stage II}: throughout the FU process, 
the servers first select some historical rounds based on a given target client and obtain the corresponding historical gradients and global models. 
Based on these selected historical gradients, global models, and assistive information uploaded during FL training, the servers jointly calculate an approximation to the required calibration to the model. Periodically, the servers check if the cumulative error surpasses a pre-defined threshold and instruct remaining clients to compute exact gradients for correction when the accumulated error is beyond the defined threshold. This process continues until all historical gradients and global models have been appropriately calibrated, indicating that the data held by the target client has been successfully unlearned. A detailed description of Starfish is presented in Algorithm \ref{alg:starfish}.

\begin{algorithm*}[h!]
\SetAlgoNoEnd
\caption{The Starfish Scheme}
\label{alg:starfish}
\KwIn{A set of $n$ federated learning clients where each client $c^j$ possesses a local dataset $D^j$, two non-colluding servers $S_1,S_2$, the initial global model $M_0$, the round number $T$ when the unlearning request is received, the total number of unlearning rounds $T^{\prime}=\lceil \sigma T \rceil$, the target client $c^t$, the learning rate $\eta_l$, the unlearning rate $\eta_u$, the selection rate $\sigma$, the tolerance rate $\alpha$, interval rate $\beta$, the buffer size $B$, the size of the global model $m$, the correction interval $T_c=\lceil \beta T \rceil$.}
\KwOut{The unlearned global model $\hat{M}_{T^{\prime}}$.}
\SetKwFunction{FL}{\textbf{Stage I. Privacy-Preserving Federated Learning}}
\SetKwProg{Fn}{}{:}{\KwRet}
\Fn{\FL}{
\For{$i \leftarrow 1$ \KwTo $T$ }{
\For{$j \leftarrow 1$ \KwTo $n$ in parallel}{
The client $c_j$ trains its local model by computing its gradients $G_i^j$ based on the global model $M_{i-1}$ and $D^j$\;
The client $c_j$ calculates the $\ell_2$-norm of its gradients $\|G_i^j\|,$ the threshold $\delta_i^j$ based on $G_i^j$ and the tolerance rate $\alpha$; \tcp{Calculating assistive information}\
The client $c_j$ secretly shares $G_i^j$, $||G_i^j||$, and $\delta_i^j$ between two servers $S_1$ and $S_2$, denoted by $[G_i^j]$, $[||G_i^j||]$ and $[\delta_i^j]$; \tcp{Secret sharing}
}
The servers aggregate the gradients from all clients using 2PC techniques to obtain $[G_i^{agg}]=\text{SecAdd}_{\mathcal{A}}[G_i^j]$\;
The servers recover $G_i^{agg} \leftarrow \text{SecRec}([G_i^{agg}])$, calculate the global model $M_i = M_{i-1}-\eta_l G_i^{agg}$\;
}
At the end of round $T$, the servers have stored $\{[G_i]\}^T$, $[\{[||G_i||]\}^T]$, $\{M_i\}^T$ and $\{[\delta^j]\}^n$\;
}
\SetKwFunction{FU}{\textbf{Stage II. Privacy-Preserving Federated Unlearning}}
\SetKwProg{Fn}{}{:}{\KwRet}
\Fn{\FU}{
Upon receiving the unlearning request, the servers initiate the generation of required materials for the 2PC process\;
\For{$j \leftarrow 1$ \KwTo $n$ in parallel}{
The servers calculate $[\delta^j]=\text{SecMax}(\{[\delta_i^j]\}^T)$ using 2PC techniques; \tcp{Threshold determination}
}

The servers calculate $T_{\delta}$ and select historical gradients with global models through $\{[G_i]\}^{T^{\prime}}, \{M_i\}^{T^{\prime}} \leftarrow \text{SecRS}(\{[G_i]\}^T$,$[\{[||G_i||]\}^T],\{M_i\}^T,M_0,T^{\prime},T_{\delta}, c^t)$; \tcp{Round selection}
\For{$i \leftarrow 1$ \KwTo $T^{\prime}$}{
\eIf{$t_i \mod T_c == 0$ \tcp{Periodical correction}}{
The servers obtain the estimated gradients and the estimated global model through $[\hat{G}_{t_i}],\hat{M}_{t_i} \leftarrow \text{SecUE}(\{[G_{t_i}]\}^{T^{\prime}},\{M_i\}^{T^{\prime}},M_0,\eta_u,B,c^t,T^{\prime})$; \tcp{Update estimation}\
The servers calculate $f_{t_i} \leftarrow \text{SecTC}([\hat{G}_{t_i}],m,\{[\delta^j]\}^{n-1},\{c^j\}^{n-1},c^t)$; \tcp{Threshold checking}\
\For{$j \leftarrow 1$ \KwTo $n$, $j \neq t$ in parallel}{
\If{$f_{t_i}^j == 1$}{
The servers instruct the client $c^j$ to calculate and upload the secret sharing of exact updates $\bar{G}_{t_i}$ between two servers $S_1,S_2$ to replace $[\hat{G}_{t_i}]$; \tcp{Error correction}
}
}
}
{
The servers obtain the estimated gradients and the estimated global model through $[\hat{G}_{t_i}],\hat{M}_{t_i} \leftarrow \text{SecUE}(\{[G_{t_i}]\}^{T^{\prime}},\{M_i\}^{T^{\prime}},M_0,\eta_u,B,c^t,T^{\prime})$; \tcp{Update estimation}
}
}
}
\KwRet{$\hat{M}_{T^{\prime}}$}
\end{algorithm*}

\section{Theoretical Analysis}
\label{sec:theoretical_analysis}

\subsection{2PC optimizations}
\label{sec:2pc_optimizations}

This section delves into a comprehensive theoretical discussion of the optimization techniques introduced in this paper, aimed at enhancing 2PC efficiency. Key focal points include (i) the replacement of secure square root with pre-computation, (ii) the selection strategy for the switching threshold $T_{\delta}$ in replacing secure division with secure multiplication, (iii) the tradeoff involved in choosing the buffer size $B$, and (iv) alternative thresholds to minimize the number of secure comparisons in the error correction process.


\textbf{Replacement of secure square root with pre-computation.}
Here we discuss the efficiency comparison between the approach used in Algorithm \ref{alg:round_selection}, which requires clients to pre-compute and send $\|g_i^t\|$ in order to avoid the need for secure calculation of norm for a given secretly shared vector which requires secure square root calculation, and an alternative where client pre-computation is not used. Note that to replace square root pre-computation, for each $i,$ instead of having $[\|g_i^t\|]$ pre-computed by the client, such calculation needs to be done in a secure manner. Recall that we aim to avoid the use of the square root computation which is required in the standard $\ell_2$-norm calculation. So to achieve this, we propose a minor modification where square root computation can be avoided. Here in order to perform the comparison, we define the following setting. Note that we are given a list of $T$ pairs $A_1\triangleq ([\mathbf{x}_1],\mathbf{y}_1),\cdots, A_T\triangleq ([\mathbf{x}_T],\mathbf{y}_t)$ where $\mathbf{x}_i = g_i^t,$ which is secretly shared and $\mathbf{y}_i = M_i - M_{i-1},$ which is a public value. The aim is to sort the list $(A_1,\cdots, A_T)$ to $(A_1',\cdots, A_T')$ of a decreasing order where $A_i\succcurlyeq A_j$ if and only if $\frac{\langle \mathbf{x}_i,\mathbf{y}_i\rangle}{\|\mathbf{x}_i\|\cdot\|\mathbf{y}_i\|}.$ 

Intuitively, the algorithm can be achieved by having each term to be replaced by its square, which prevents the need of the square root computation. However, this needs to be done carefully since $\langle \mathbf{x}_i,\mathbf{y}_i\rangle$ can be negative in which comparison may no longer be correct when the values are squared. In order to preserve the sign of each inner product, we introduce a new variable to store the sign of $\langle \mathbf{x}_i,\mathbf{y}_i\rangle$ and $\text{SecGE3}$ and $\text{SecGE4},$ which takes the sign variable into account during comparison calculation. The complete discussion of $\text{SecRS}^{(alt)}$ can be found in Appendix \ref{app:SRS}.


\textbf{Switching threshold $\bm{T_{\delta}}$.}
We determine the value of $T_{\delta}$ for Algorithm \ref{alg:round_selection} under the assumption that $\text{SecDiv}$ is realized using ABY \cite{demmler2015aby,PSSY21} and the sorting process is done using the protocol proposed in \cite{Bat68}. Given this, we consider the additional complexity incurred by Methods $1$ and $2$ compared to each other.
\begin{enumerate}
    \item \textit{Method 1}: Note that a call to $\text{SecMul}$ requires $2\ell_q$ bits of communication and the secure sorting used in ABY is the bitonic sort \cite{FNO22} proposed in \cite{Bat68}, which requires $\frac{\ell_T^2+\ell_T}{4} T$ calls to the underlying comparison protocol for a list of length $T$ where $\ell_T=\lceil\log T\rceil.$ Hence, since we require $2$ calls of $\text{Secmul}$ for each call to $\text{SecGE2},$ the additional bits of communication incurred by Method $1$ which does not exist in Method $2$ is $\ell_q\left(\ell_T^2+\ell_T\right)T.$
    \item \textit{Method 2}: Here we assume that the implementation of $\text{SecDiv}$ in ABY is equivalent to the secure remainder calculation proposed in the modular exponentiation protocol \cite{demmler2015aby}, which is done by first converting the shares to Yao-type share. In such case, $\text{SecDiv}$ requires $10\ell_q^2 + \ell_q(\kappa+1)$ bits of communication where $\ell_q=\lceil \log q\rceil$ and $\kappa$ is the security parameter. This implies that using the second approach to find the $T'$ largest pairs requires additional $T\ell_q(10\ell_q + \kappa+1)$ bits of communication which does not exist in Method $1.$
\end{enumerate}

Comparing the two extra bits of communications from the two proposed methods, it is then easy to see that Method $1$ requires less communication complexity if $T<2^{\frac{\sqrt{1+40\ell_q+4(\kappa+1)}-1}{2}}.$



\textbf{Buffer size $\bm{B}$.} As described earlier, derived from \cite{byrd1994representations}, we use a recent set $B$ of $\Delta G=\{[\Delta G_{t_{i}}]\}^B$ and $\Delta M=\{\Delta M_{t_{i}}\}^B$ in a buffer, rather than those from all rounds, to approximate $\mathcal{H}_{t_{i}}^{-1}$. This incurs an additional storage of $Bm(n-1)(\ell_q+\ell)$ bits. We first note that in \text{SecUE}, the iterations of $t_i$ need to be done sequentially since we need the computation result of the previous round to perform the calculation of the next round. However, in Algorithm \ref{alg:update_estimation}, the $n-1$ iterations for the value $j$ can be concurrently executed. This implies that such a loop can be completed in the same number of rounds as any single iteration, although the communication complexity will be multiplied by $n-1.$ Lastly, we note that the only functionalities that require any communication in Algorithm \ref{alg:update_estimation} are $\text{SecMI}, \text{SecMul3}, \text{SecMul}$ and $\text{SecRec}$ with $m\times m$ matrices as inputs respectively. According to \cite{PSSY21}, we can see that $\text{SecMul3}$, $\text{SecMul}$, and $\text{SecRec}$ each require one communication round with $2m^2\ell_q$ bits of communication. On the other hand, by Algorithm~\ref{alg:matrix_inversion}, $\text{SecMI}$ requires two $\text{SecMul}$ calls and one $\text{SecRec}$ call. Hence, in total it takes $3$ communication rounds with $6m^2\ell_q$ bits of communication required.

Simultaneously, the recursive computation of $\mathcal{H}_{t{i}}^{-1}$ based on the stored $\Delta G$ and $\Delta M$ introduces additional 2PC operations with $7B$ communication rounds with $ (14n-12) m^2\ell_q B$ bits being communicated in total. We can observe that a larger $B$ results in an increase in storage requirements as well as a larger communication overhead for the 2PC operations. However, as we will observe in Theorem \ref{theo_1}, it provides a smaller upper bound in conjunction with other given parameters for the difference between the resulting model and the trained-from-scratch model.

\textbf{Alternative thresholds.} In Section \ref{sec:correcting_errors}, we demonstrate that the straightforward adaptation of the threshold determination from FedRecover \cite{cao2023fedrecover} to a privacy-preserving version necessitates the call of $\text{SecSrt}$ over all stored historical gradients, totalling to $mnT$ entries. So assuming bitonic sort is used, this requires $\frac{1}{2}\ell_{mnT} (\ell_{mnT}+1)$ rounds of parallel calls of secure comparison totalling to $\frac{\ell_{mnT}^2 + \ell_{mnT}}{4} mnT$ calls to secure comparison protocol where $\ell_{mnT}\triangleq \lceil \log mnT\rceil.$
In contrast, in the Starfish scheme, we set different thresholds for different clients. Hence, the servers must securely compute a threshold for each client in every round. Subsequently, the thresholds from all historical rounds are collected, allowing the servers to securely identify the maximum value. In this case, the servers perform SecSrt over $m$ entries $T(n-1)$ times in parallel, followed by performing SecMax over $T$ entries $n-1$ times in parallel. 
Letting $\ell_m=\lceil\log m\rceil,$ the sorting step requires $\frac{1}{2}\ell_m (\ell_m +1)$ rounds of parallel calls of secure comparison totalling to $\frac{\ell_{m}^2 + \ell_{m}}{4} mT(n-1)$ calls of the secure comparison protocol. On the other hand, each call of $\text{SecMax}$ over $T$ entries requires $\ell_T\triangleq \lceil \log T\rceil$ rounds of parallel calls to secure comparison protocol totaling up to $T-1$ calls of secure comparison protocol. Hence in total, this second phase requires $\ell_T$ rounds of parallel calls to secure comparison protocol totaling up to $(T-1)(n-1)$ calls of secure comparison protocol.
Hence, in total, our approach requires $\frac{1}{2}\ell_m(\ell_m+1) + \ell_T$ rounds of parallel calls to secure comparison protocol totaling up to $ \frac{\ell_{m}^2 + \ell_{m}}{4} mT(n-1) + (T-1)(n-1)$ calls of secure comparison protocol. It can be easily observed that our approach provides an improvement in both the number of rounds of parallel calls of secure comparison protocol as well as the total number of calls to such protocol.

\subsection{Resource consumption}
In this section, we provide analyses of resource consumption in terms of memory and communication. Note that since 2PC techniques are utilized for privacy preservation purposes, the communication overhead dominates the overall cost, compared to the computation overhead. Hence, we only provide an analysis of memory and communication. A summary of these analyses can be found in Appendix \ref{app:RouInComp}.

\textbf{Memory.} The memory consumption in the Starfish scheme is solely managed on the server-side and stems from (i) historical gradients and global models, (ii) assistive information, (iii) pre-computed materials for 2PC operations, and (iv) intermediate results stored in the buffer for Hessian approximation.

For historical gradients and global models, upon receiving the unlearning request at round $T$, each server has already incurred a memory consumption of $Tmn(\ell_q+\ell)$ bits, where $\ell$ is the bit length of plaintext, and $\ell_q$ is the bit length of an element in $\mathbb{F}_q$. This is because the servers store gradients over secret shares and global models over plaintext. The storage of assistive information, including thresholds and $\ell_2$-norm, incurs additional memory consumption of $2Tn\ell_q$ bits. The intermediate results stored in the buffer result in a memory consumption of $Bm(n-1)(\ell_q+\ell)$ bits, as analyzed in Section \ref{sec:2pc_optimizations}.

By the round and invocation complexity analyses of the Starfish scheme, which can be found in Appendix \ref{app:RouInComp}, the total memory for the auxiliary values generated during the offline phase is $O(\kappa \log q \log \log q \max(nT,T\log^2T,nm) + T'B(n-1)m^2\log q )$ bits.


\textbf{Communication.}
We will now discuss the communication overheads on both the client-side and server-side.

The client-side communication cost includes (i) the secret sharing of gradients with assistive information between two servers during FL training and (ii) the secret sharing of gradients between servers for periodic correction in the unlearning process. For the communication overhead during FL training, each client is required to share its gradient and assistive information with a communication complexity of $T(m+2) \ell_q$ bits. During the periodic correction in the unlearning process, each client computes and secretly shares its gradients between two servers with a communication complexity of $\lfloor \beta^{-1} \rfloor m \ell_q$ bits.
Due to the page limitation, the complexity analyses of our proposed scheme can be found in Appendix \ref{app:RouInComp}.

\subsection{Bounding the difference}
We demonstrate that the difference between the unlearned global model acquired through Starfish and the one obtained from a train-from-scratch approach can be quantifiably bounded, given certain assumptions, which are elaborated in Appendix \ref{app:ModDif}.

\begin{theorem}[Model difference between Starfish and train-from-scratch]\label{theo_1}
    The difference between the unlearned global model obtained through Starfish at $t_i$ and the global model obtained through train-from-scratch at $t=\lceil \frac{1}{\sigma}t_i\rceil$ can be bounded as follows:
    \begin{equation}
        \Vert \hat{M}_{t_i}-M_t \Vert 
        \leq 2\sqrt{ \eta_u [\frac{1}{\mu} + \frac{1}{\sigma(\mu-2)}][F({M}_{0})-F({M^*})] t_i  }
\end{equation}
where $\eta_u$ is the unlearning rate in the unlearning progress, $M_0$ is the initial model used in both Starfish and train-from-scratch, $M^*$ is the optimal solution for the objective function $F(M)$ with $\mu$-strongly convex and $L$-smooth.
\end{theorem}

Referencing Theorem \ref{theo_1}, the proof of which is detailed in Appendix \ref{app:ModDif}, we establish an upper bound for the difference between global models derived from the Starfish and train-from-scratch approaches. In Starfish, the round selection is guided by the selection rate $\sigma$, where a higher $\sigma$ leads to a smaller difference bound. This presents a trade-off between storage costs and model accuracy, as selecting more historical models improves the accuracy of the unlearned global model.


\begin{figure}[htb]
    \centering
    \begin{subfigure}{0.48\linewidth}
        \includegraphics[width=\linewidth]{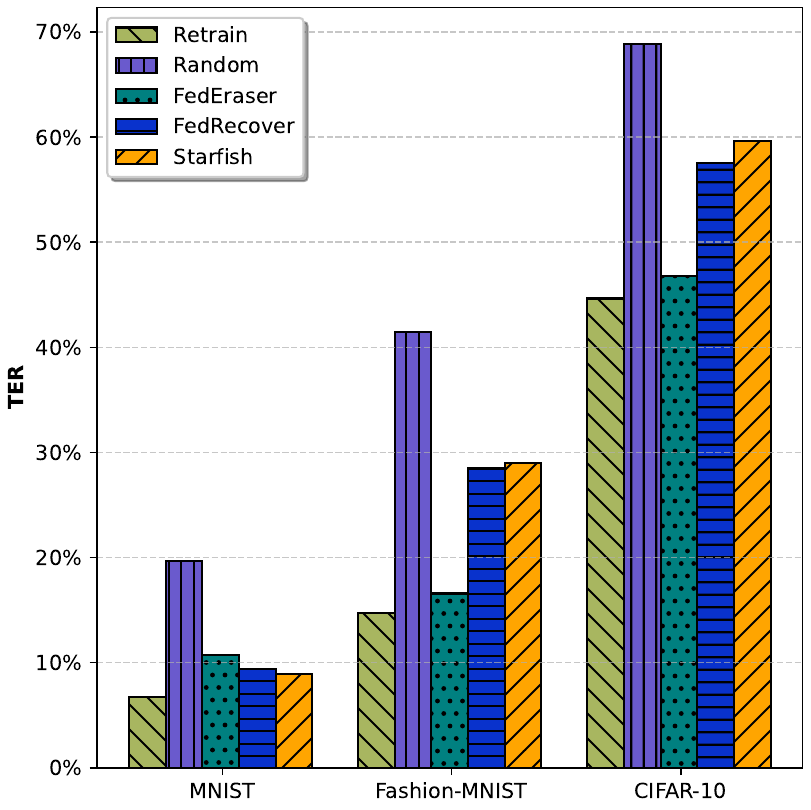}
        \caption{TER}
        \label{fig:ter}
    \end{subfigure}
    \begin{subfigure}{0.48\linewidth}
        \includegraphics[width=\linewidth]{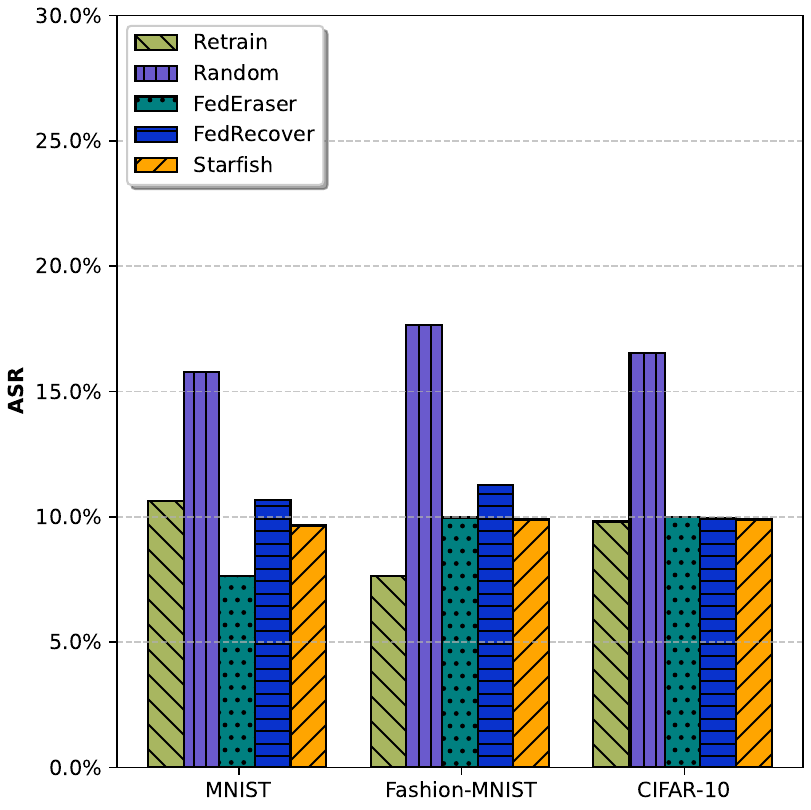}
        \caption{BA-ASR}
        \label{fig:asr_ba}
    \end{subfigure}
    
    \medskip
    
    \begin{subfigure}{0.48\linewidth}
        \includegraphics[width=\linewidth]{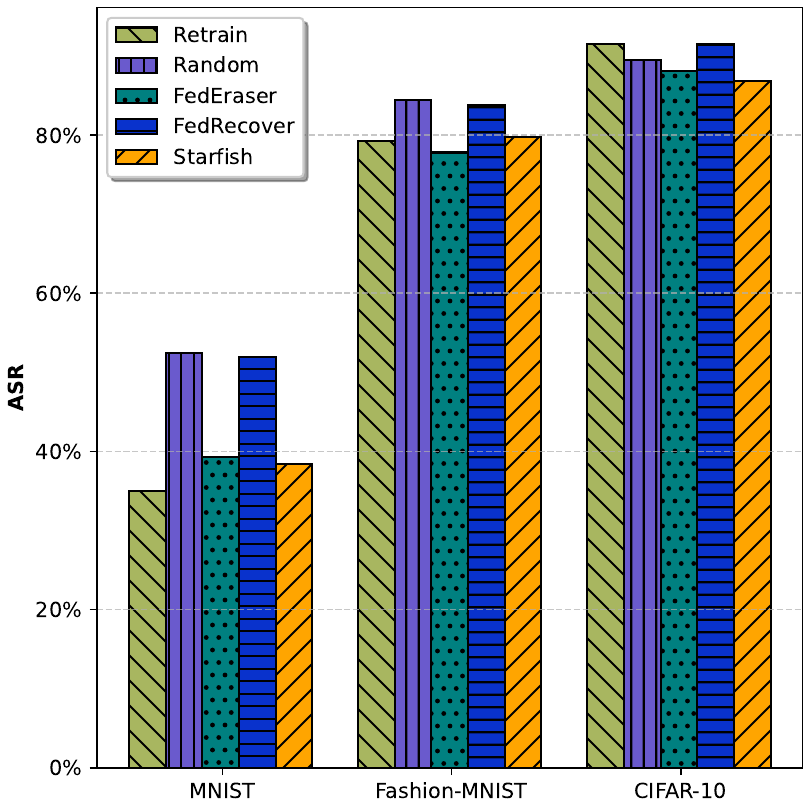}
        \caption{MIA-ASR}
        \label{fig:asr_mia}
    \end{subfigure}
    \begin{subfigure}{0.48\linewidth}
        \includegraphics[width=\linewidth]{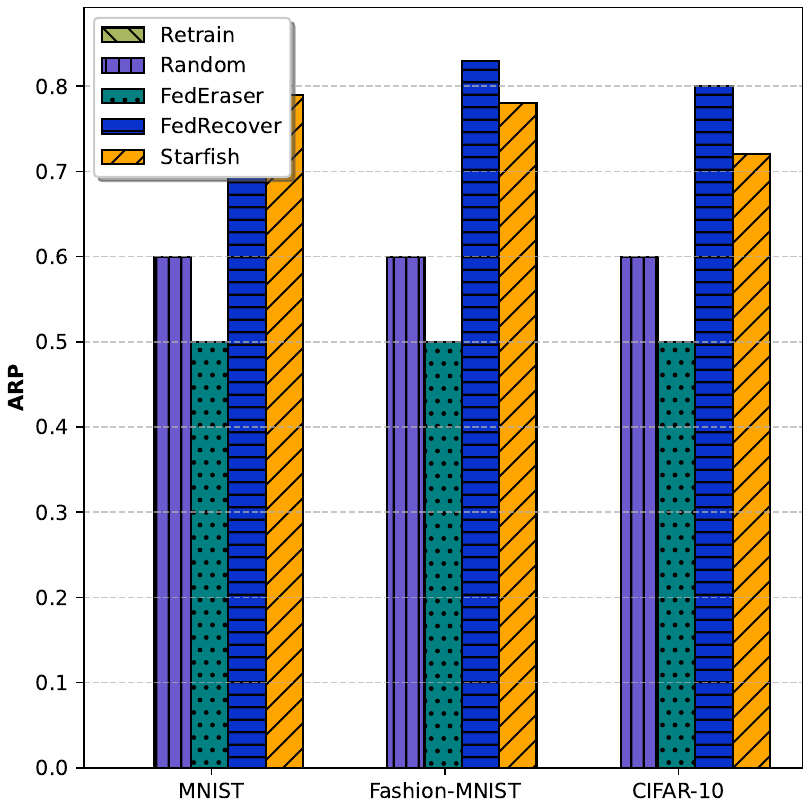}
        \caption{ARP}
        \label{fig:arp}
    \end{subfigure}
    \caption{Comparison of Test Error Rate (TER) on the test dataset, Attack Success Rate over Backdoor Attacks (BA-ASR), Membership Inference Attacks (MIA-ASR), and Average Round-saving Percentage (ARP) across three ML tasks with varying complexity. Smaller TER and ASR indicate better accuracy, while a larger ARP implies enhanced efficiency.}
    \label{fig:ter_asr_arp}
\end{figure}

\section{Experimental Evaluation}
\label{sec:experimental_valuation}

\subsection{Experiment settings.}
\textbf{Unlearning setting.} In the Starfish scheme, unless otherwise specified, the selection rate $\sigma$ is set to 0.6, the tolerance rate $\alpha$ to 0.4, and the interval rate $\beta$ to 0.1. 
Our approach for the unlearning process aligns with methods outlined in \cite{liu2021federaser,cao2023fedrecover}, employing FedAvg \cite{mcmahan2017communication} as our aggregation method, involving 20 clients, each conducting 5 local training rounds, cumulatively resulting in 40 global epochs. Stochastic Gradient Descent (SGD) without momentum is used as the optimization strategy for the clients, with a uniform learning rate of 0.005 and a batch size of 64.

\textbf{2PC setting.} In the 2PC protocol, we tested our scheme on a machine powered by an Intel(R) Xeon(R) W-2123 CPU @ 3.60GHz with 16GB of RAM, running Ubuntu 22.04.2 LTS. For the implementation, we implement our schemes in C++, utilizing the ABY framework \cite{demmler2015aby}. In the context of fixed-point arithmetic, all numerical values are represented within the domain of $\mathbb{Z}_{2^{64}}$, with the lowest 13 bits designated for the fractional component. The chosen security parameter is 128 bit. We simulate the network environment connection using the Linux \textit{trickle} command. For the LAN network, we set the upload and download bandwidth to 1GB/s and network latency to 0.17ms.  For the experiment to simulate a WAN network, we limit the upload and download bandwidth to 100MB/s, and we set the network latency to 72ms. We averaged our experiment results for 10 runs.

\textbf{Baselines.} We compare the Starfish scheme with three baseline methods:

\begin{itemize}
    \item Train-from-scratch: In this baseline method, we remove the target client and then follow the standard federated learning process to retrain a global model from scratch with the remaining clients.
    \item Random round selection: Given a selection rate $\sigma$, distinct from the cosine similarity-based round selection, the servers randomly choose $T^{\prime}$ rounds, upon which they initiate the unlearning process.
    \item FedEraser \cite{liu2021federaser}: As mentioned earlier, FedEraser shares a similar design structure with the Starfish scheme, utilizing historical gradients and global models for calibrations conducted by remaining clients to unlearn the data of the target client.
    \item FedRecover \cite{cao2023fedrecover}: FedRecover minimizes computation overhead on the client-side by instructing the server to estimate client updates for calibration based on historical gradients and global models. Periodic corrections conducted by the remaining clients are also employed to mitigate errors resulting from the estimation process.
\end{itemize}

\textbf{Evaluation metrics.} To evaluate unlearning performance, we employ several metrics and implement verification methods based on Backdoor Attacks (BA) and Membership Inference Attacks (MIA). We outline them as follows:


\begin{itemize}
    \item Test Error Rate (TER): TER represents the proportion of test inputs that the global model incorrectly predicts.
    \item BA Attack Success Rate (BA-ASR): BA-ASR is the fraction of target clients' data predicted to have the target label with the backdoor trigger.
    \item MIA Attack Success Rate (MIA-ASR): MIA-ASR is defined as the fraction of the target clients' data that is predicted to have membership through MIA inference.
    \item Average Round-saving Percentage (ARP): Denoting $T_r$ as the number of rounds that the client is instructed to participate in error correction, ARP is defined as the average percentage of round-saving for the clients, calculated as $(T - T_r) / T \times 100\%$.
\end{itemize}

Additionally, we evaluate the 2PC performance by assessing communication costs and runtime for each step and overall, considering different Starfish parameter configurations.

\textbf{Machine learning tasks.} We assess the Starfish scheme across neural networks with diverse sizes and widths, employing configurations referenced in \cite{cao2023fedrecover}. For MNIST, we utilize two CNN layers and two fully connected layers. In the case of Fashion-MNIST (F-MNIST), we add an extra fully connected layer. For CIFAR-10, we increase the width of the fully connected layers from 1024 to 1600 dimensions.


\subsection{Experimental Evaluations.}

In this section, we present the experimental results and discuss them from two perspectives, focusing on unlearning performance and 2PC performance.

\subsubsection{Unlearning performance.}

We now discuss the unlearning performance of Starfish.

\textbf{Unlearning effectiveness.} Figure \ref{fig:ter_asr_arp} compares the unlearning performance of Starfish with four baselines, using four evaluation metrics across three datasets. We have observed that Starfish attains TERs and ASRs comparable to those of FedRecover. It also exhibits slightly higher ASRs for both backdoor and membership inference attacks compared to the train-from-scratch approach. However, these ASRs are lower than those achieved using the random round selection method and FedEraser. Additionally, Starfish demonstrates a lower ARP relative to FedRecover but outperforms other baseline methods. This suggests an inherent trade-off, wherein Starfish incurs additional costs for enhanced privacy guarantees compared to FedRecover.

\begin{figure}[htbp!]
    \centering
    \begin{subfigure}{0.48\linewidth}
        \centering
        \includegraphics[width=0.9\linewidth]{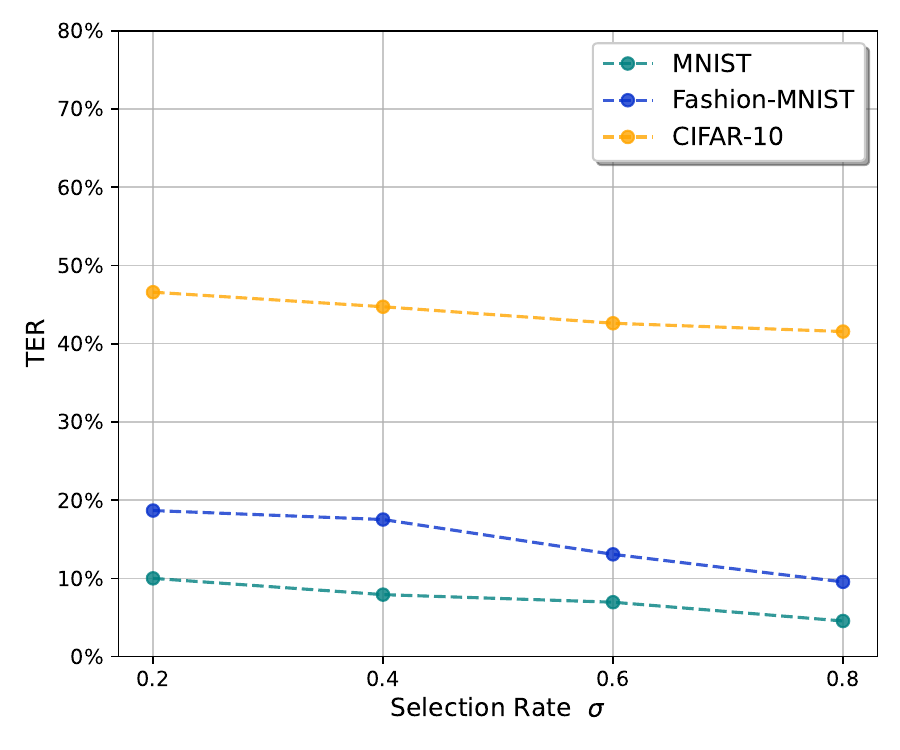}
        \caption{TER}
        \label{fig:sel_rt_ter}
    \end{subfigure}
    \begin{subfigure}{0.48\linewidth}
        \centering
        \includegraphics[width=0.9\linewidth]{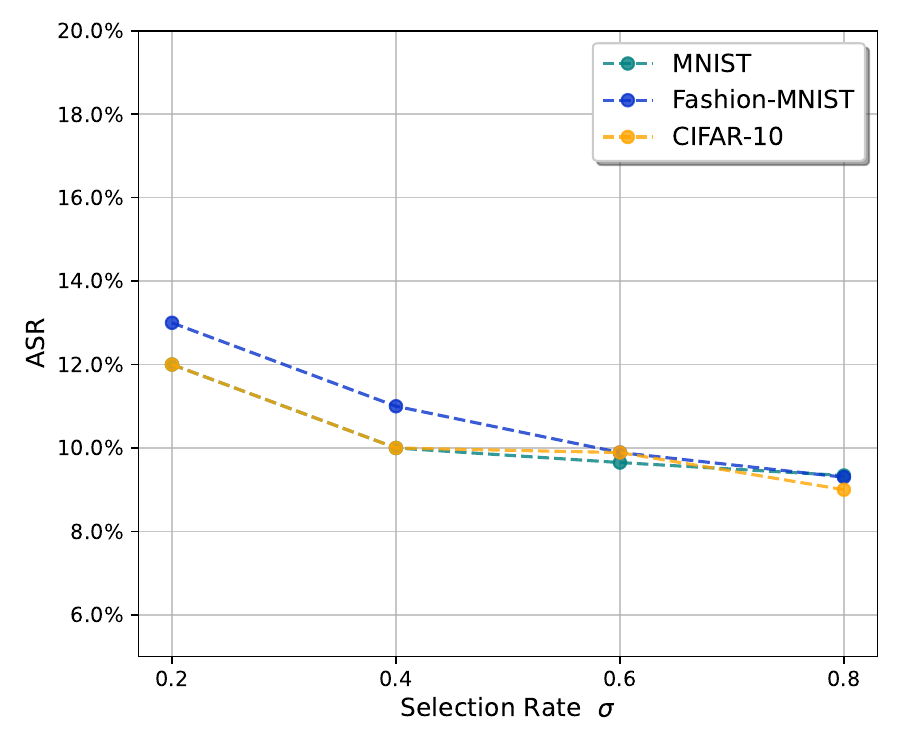}
        \caption{ASR}
        \label{fig:sel_rt_asr}
    \end{subfigure}

    \caption{Impact of the selection rate $\sigma$ on the unlearning performance of the Starfish scheme.}
    \label{fig:sel_rt_impact}
\end{figure}

\textbf{Impact of the Starfish parameter configuration.} An example is provided in Figure \ref{fig:sel_rt_impact} to illustrate the impacts of the selection rate $\sigma$ on the unlearning performance of Starfish respectively. These impacts are evaluated using TERs and BA-ASRs. A consistent increase in TERs is observed as the complexity of the underlying ML tasks increases. Additionally, both TERs and ASRs rise with a higher selection rate $\sigma$. The reason is that a larger selection rate reduces the use of historical information, which aligns with the intended design of the Starfish scheme. Similar observations on the tolerance rate $\alpha$ and interval rate $\beta$ can be obtained from the additional experimental results given in Appendix \ref{sec:add_exp}.


\begin{figure}[htbp!]
    \centering
    \begin{subfigure}{0.48\linewidth}
        \centering
        \includegraphics[width=1\linewidth]{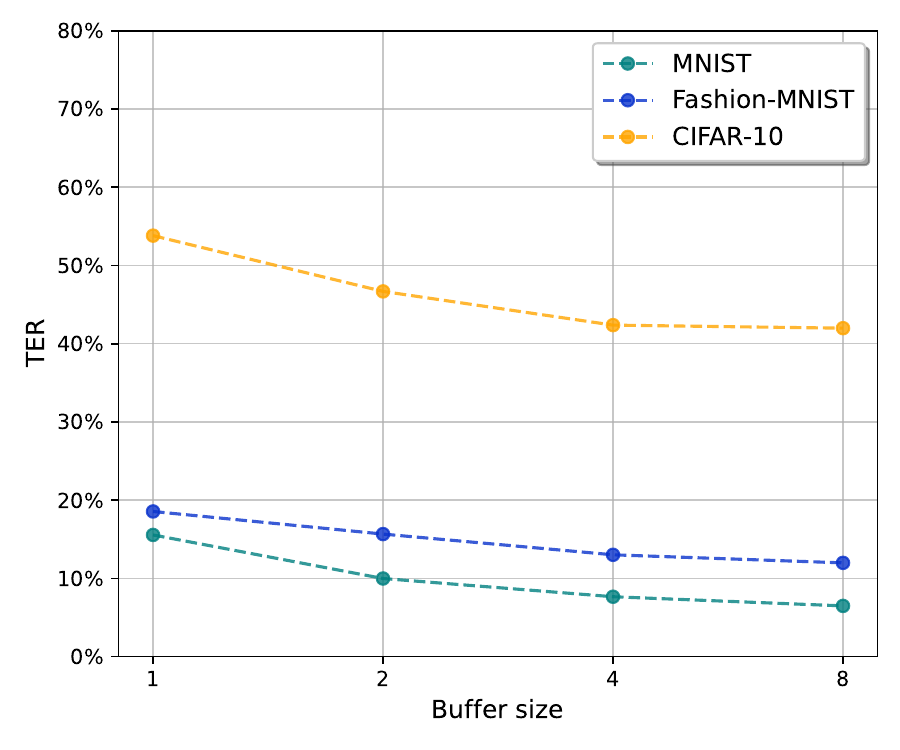}
        \caption{TER}
    \end{subfigure}
    \begin{subfigure}{0.48\linewidth}
        \centering
        \includegraphics[width=1\linewidth]{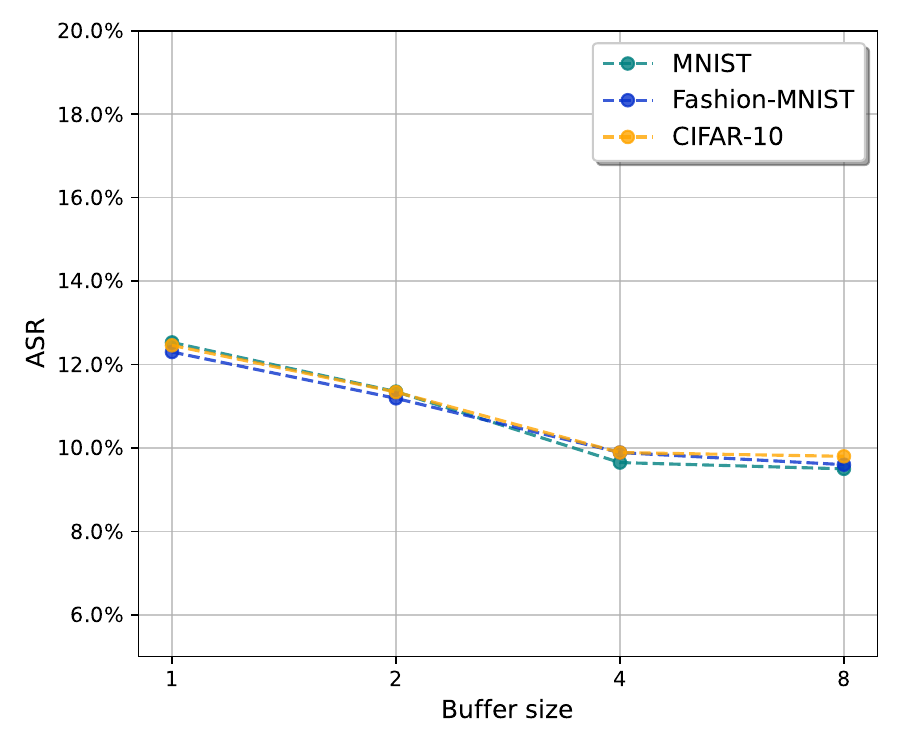}
        \caption{ASR}
    \end{subfigure}
    \caption{Impact of the buffer size $B$ on the unlearning performance of the Starfish scheme.}
    \label{fig:int_rt_impact}
\end{figure}

\begin{figure}[ht!]
    \centering
    \begin{subfigure}{0.23\textwidth}
        \includegraphics[width=1\linewidth]{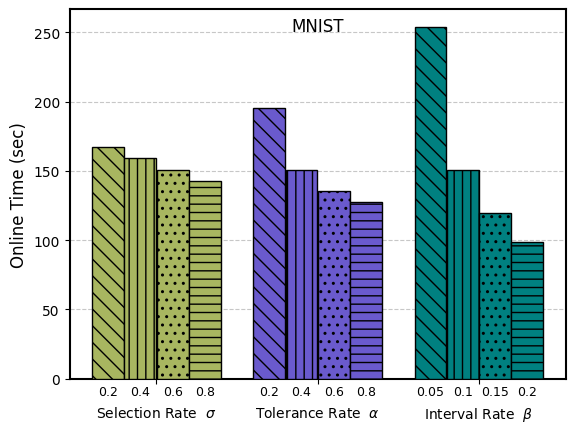}
        \caption{Online time}
        \label{fig:online_time_MNIST}
    \end{subfigure}
    \begin{subfigure}{0.225\textwidth}
        \includegraphics[width=1\linewidth]{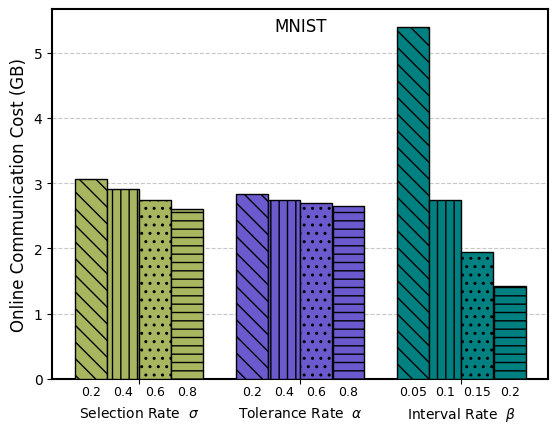}
        \caption{Online communication}
        \label{fig:online_comm_MNIST}
    \end{subfigure}

    \caption{Comparison of different parameter configurations on MNIST dataset.}
    \label{fig:para_time_comm_MNIST}
\end{figure}

\begin{figure}[ht!]
    \centering
    \begin{subfigure}{0.23\textwidth}
        \includegraphics[width=1\linewidth]{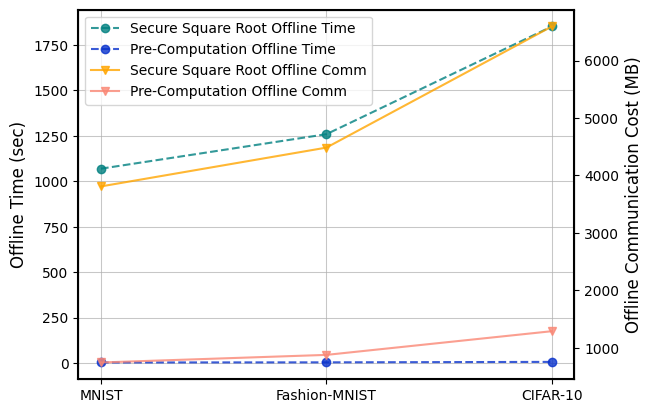}
        \caption{Offline time and comm}
        \label{fig:square_root_offline}
    \end{subfigure}
    \begin{subfigure}{0.23\textwidth}
        \includegraphics[width=1\linewidth]{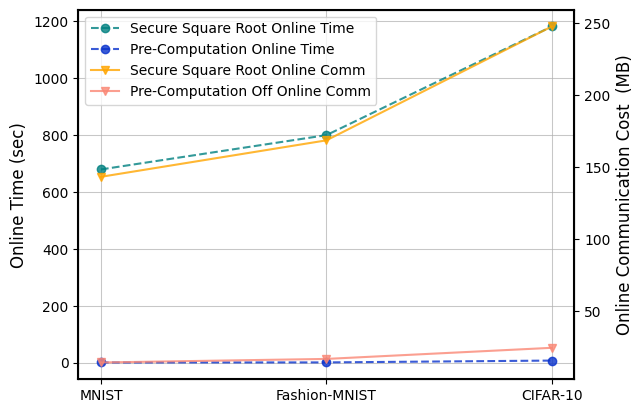}
        \caption{Online time and comm}
        \label{fig:online_comm}
    \end{subfigure}

    \caption{Comparison of the secure square root method and the pre-computation method.}
    \label{fig:square_root}
\end{figure}
\subsubsection{2PC performance.}
We now present the 2PC performance of the Starfish scheme, evaluating it in terms of communication cost and runtime. Additionally, we examine how various parameter configurations impact the 2PC efficiency. Unless otherwise specified, the 2PC experiments are conducted over a single unlearning round (line 15-24 in Algorithm \ref{alg:starfish}). As outlined in Section \ref{sec:2pc_optimizations}, we initially evaluate the 2PC performance of specific optimizations in the Starfish scheme, followed by an analysis of the overall cost associated with the Starfish scheme. A summary of the 2PC performance of the unlearning process in Starfish, evaluated in both LAN and WAN settings, is provided in Table \ref{tab:time_comm}.

\textbf{Replacement of secure square root.} 
Figure \ref{fig:square_root} compares the original 2PC-based FedRecover, using secure square root, with Starfish's pre-computation method. Starfish shows improved efficiency in online evaluation, despite a small increase in offline time and communication costs. This efficiency gain is more significant for complex ML tasks, highlighting Starfish's scalability in advanced ML applications.

\textbf{Threshold determination and round selection.} The comparison between the 2PC implementation of threshold determination in FedRecover ($\text{TD}^1$) and in Starfish ($\text{TD}^2$), as well as the round selection based on Method 1 ($\text{RS}^1$) and Method 2 ($\text{RS}^2$), is detailed in Table \ref{tab:step_one_round}. Note that results provided in Table \ref{tab:step_one_round} are obtained from ML tasks using the MNIST dataset. Results involving ML tasks of different complexities can be found in Appendix \ref{sec:add_exp}.

\textbf{Buffer size $B$.} 
As illustrated in Figure \ref{fig:int_rt_impact}, an increased buffer size $B$ enhances the performance of unlearning. However, this improvement comes at the expense of higher communication costs and extended runtime, as further detailed in Appendix \ref{sec:add_exp}. Therefore, a clear trade-off emerges. FedRecover \cite{cao2023fedrecover} sets the buffer size $B$ to 2 to achieve an optimal balance.

\textbf{Impact of the Starfish parameter configuration.} 
Figure \ref{fig:para_time_comm_MNIST} demonstrates that increasing the selection rate $\sigma$ and tolerance rate $\alpha$ reduces both online time and communication costs per unlearning round in the Starfish scheme. However, raising the interval rate $\beta$ decreases these costs but raises the total unlearning process cost. This pattern, confirmed by Figure \ref{fig:sel_rt_impact} and Appendix \ref{sec:add_exp}, reveals a trade-off between unlearning efficiency and 2PC performance. Higher unlearning efficiency, via higher $\sigma$ and $\alpha$ and lower $\beta$, leads to greater communication costs and runtime, and vice versa.

\begin{table}[htb]
\small
\caption{Comparison of 2PC performance for each step in a single unlearning round on MNIST dataset.}
\label{tab:step_one_round}
\centering
\begin{tabular}{lcccc}
\hline \\[-1em]
Step & \multicolumn{1}{c}{\begin{tabular}[c]{@{}c@{}}Offline\\ Time {\scriptsize(s)}\end{tabular}} & \multicolumn{1}{c}{\begin{tabular}[c]{@{}c@{}}Online\\ Time {\scriptsize(s)}\end{tabular}} & \multicolumn{1}{c}{\begin{tabular}[c]{@{}c@{}}Offline\\ Comm {\scriptsize(GB)}\end{tabular}} & \multicolumn{1}{c}{\begin{tabular}[c]{@{}c@{}}Online\\ Comm {\scriptsize(GB)}\end{tabular}}  \\ \hline \\[-1em]
$\text{TD}^1$ & 0.04  & 0.04  & 4.20  & 0.12 \\
$\text{TD}^2$ & 0.01  & 0.02  & 0.11  & 0.01 \\
$\text{RS}^1$ & 145.1 & 21.80 & 29.80 & 0.56 \\
$\text{RS}^2$ & 87.15 & 13.14 & 17.88 & 0.34 \\
$\text{UE}$  & 3.59   & 1.68  & 2.14  & 0.04 \\ 
$\text{TC}$  & 689.44 & 103.76 & 141.55 & 2.66 \\\hline \\[-1em]
\end{tabular}
\caption*{{\scriptsize
  $\text{TD}^1$: Threshold Determination in FedRecover;
  $\text{TD}^2$: Threshold Determination in Starfish;
  $\text{RS}^1$: Round Selection of \textit{Method 1};
  $\text{RS}^2$: Round Selection of \textit{Method 2};
  $\text{UE}$: Update Estimation;
  $\text{TC}$: Threshold Checking.
}}
\end{table}

\begin{table}[htb]
\small
\caption{2PC performance comparison across various datasets under different network settings on MNIST dataset.}
\label{tab:time_comm}
\centering
\begin{tabular}{ccccc}
\hline \\[-1em]
& \multicolumn{1}{c}{\begin{tabular}[c]{@{}c@{}}Offline\\ Time {\scriptsize(s)}\end{tabular}} 
& \multicolumn{1}{c}{\begin{tabular}[c]{@{}c@{}}Online\\ Time {\scriptsize(s)}\end{tabular}} 
& \multicolumn{1}{c}{\begin{tabular}[c]{@{}c@{}}Offline\\ Comm {\scriptsize(GB)}\end{tabular}} 
& \multicolumn{1}{c}{\begin{tabular}[c]{@{}c@{}}Online\\ Comm {\scriptsize(GB)}\end{tabular}} 
\\ \hline \\[-1em]
LAN & 18826.21 & 3623.81 & 3530.85 & 66.27 \\
WAN & 25801.17 & 3735.61 & 3562.54 & 66.93 \\ \hline \\[-1em]
\end{tabular}
\end{table}





\section{Conclusions}
\label{sec:conclusions}
In summary, our Starfish scheme pioneers a privacy-preserving federated unlearning approach, leveraging 2PC techniques and tailored optimization upon state-of-the-art FU method. The provided theoretical bound underscores its efficacy, showing minimal differences between the unlearned global model obtained through Starfish and one derived from train-from-scratch for certified client removal. Empirical results validate Starfish's efficiency and effectiveness, affirming its potential for privacy-preserving federated unlearning with manageable efficiency overheads.

\bibliographystyle{plain}
\bibliography{mybibliography}

\appendix

\section{Notations.}
\label{sec:notations}
We summarize the parameters and notations used throughout the paper in Table \ref{tab:notations}.

\begin{table}[htb]
\centering
\caption{Notations of parameters used throughout the paper.}
\label{tab:notations}
\begin{tabular}{c|l}
\toprule
Notation & Description \\ \midrule
$n$ & Number of participating clients  \\
$m$ & The size of the global model  \\
$\ell$ & The bit length of plaintext \\
$\ell_q$ & The bit length of an element in $\mathbb{F}_q$ \\
$i$ & Round index  \\
$j$ & Client index  \\
$c^t$ & The target client  \\
$M_0$ & The initial global model  \\
$M_i$ & The global model in round $i$  \\
$T$ & Total number of learning rounds  \\
$T^{\prime}$ & Total number of unlearning rounds  \\
$T_{\delta}$ & Switching threshold \\
$G_i$ & Historical gradients in round $i$  \\
$\hat{G}_i$ & Estimated gradients in round $i$  \\
$\bar{G}_i$ & Corrected gradients in round $i$  \\
$\delta^j$ & The threshold for client $c^j$  \\
$T_c$ & Correction interval  \\
$\alpha$ & Tolerance rate  \\
$\beta$ & Interval rate  \\
$\sigma$ & Selection rate  \\
$B$ & The buffer size  \\
\bottomrule
\end{tabular}
\end{table}

\section{Secure two-party computation}\label{app:2pc}
\subsection{Fixed-Point Arithmetic}\label{app:2pc-fpa}
First, we briefly discuss the general idea of fixed-point encoding and define some relevant notations. First, we assume the existence of a positive integer $e$ such that for any value $x$ that is considered, $x \in \mathbb{R}$ and $|x|<2^{e-1}.$ We further assume that the participants agree on a precision parameter $p\in \mathbb{Z}_{>0}$ which means that any $x\in \mathbb{R}$ is stored with an accuracy of at most $2^{-p}.$ This implies the existence of a set $S_{e,p}\subseteq \mathbb{R}$ where for any value $y\in S_{e,p},$ there exist $d_{-p},\cdots, d_{e-2}\in \{0,1\}$ such that $y=\pm \sum_{i=-p}^{e-2} d_i 2^{i}.$ Hence, given $x\in \mathbb{R},$ there exists $\tilde{x}\in S_{e,p}$ that is the closest to $x,$ namely $\tilde{x} = 2^{-p}\lfloor x\cdot 2^p\rceil.$ The encoding process goes as follows. First, we assume that $q$ is an odd prime that is larger than $2^{2p+2e+\kappa}$ for some security parameter $\kappa$ and let $\mathbb{F}_q=\left\{-\frac{q-1}{2},\cdots, -1,0,1,\cdots, \frac{q-1}{2}\right\}.$ Given $x\in \mathbb{R}, \varphi_q(x)=\bar{x}$ where $\bar{x}\equiv \lfloor2^p\tilde{x}\rceil\pmod{q}.$ Note that for such an encoding process, a truncation protocol is required for any invocation of multiplication. More specifically, suppose that $\bar{x}=\varphi_q(x),\bar{y}=\varphi_q(y)\in \mathbb{F}_q$ for some $x,y\in \mathbb{R}.$ Suppose that $2^p x = \bar{x} + r_x$ and $2^p y = \bar{y} + r_y$ for some $r_x,r_y\in \left[-\frac{1}{2},\frac{1}{2}\right).$ It is then easy to see that $x=2^{-p}\bar{x} + s_x$ and $y=2^{-p}\bar{y} + s_y$ for some $s_x,s_y\in \left[-2^{-(1+p)}, 2^{-(1+p)}\right].$ Then $xy = 2^{-2p}\bar{x}\bar{y} + 2^{-p} (s_x\bar{y} + s_y\bar{x}) + s_xs_y.$ It is then easy to see that to estimate $\bar{xy} =\lfloor 2^p xy\rceil,$ we may use $2^{-p} \bar{x}\bar{y}.$ Here the truncation protocol is used since after the multiplication by $2^{-p},$ our value will have a higher precision of $2^{-2p}.$ Hence the $p$ least significant bits need to be truncated to avoid the need of increasing precision. This can be done by simply removing the $p$ least significant bits from the product after the calculation. Note that doing so locally to shares for values stored as secret shares introduces some probability of precision error. It was shown (see for example \cite[Theorem $1$]{MZ17}) that the error size is at most as large as the precision of the least significant bit of the value except with some negligible probability.
\subsection{2PC Functionalities}\label{app:2pc-func}

In this, we briefly discuss some functionalities that we require for our underlying secret-sharing based computation.

\begin{enumerate}
    \item First, we consider some basic functionalities present in any secret-sharing based two-party computation schemes (2PC scheme for short). We assume the existence of secure protocols $\text{SecShare},\text{SecRec},$ which, transforms $v\in \mathbb{F}_q$ to $[v]$ and vice versa. That is, the two protocols generate the share for a given value $v$ and recover the original value given the shares respectively. Next, we assume the existence of a protocol $\text{SecRanGen}$ which securely generates a secretly shared random value $[v]$ where $v\in \mathbb{F}_q$ is unknown to any of the servers. We also assume the existence of secure protocols $\text{SecAdd}$ and $\text{SecMul},$ which given $[u]$ and $[v],$ returns $[u+v]$ and $[uv]$ respectively. We note that due to the linearity of additive secret sharing scheme, $\text{SecAdd}$ can be done without any communication between the servers. We also note that local computation can also be done if $\text{SecAdd}$ or $\text{SecMul}$ receive inputs with at least one of the values being in the clear. Lastly, $\text{SecMul}$ with two secretly shared inputs requires communication between the servers. It can be achieved, for instance, by the use of Beaver triple \cite{Bea92}. Note that if Beaver triple is used in realizing $\text{SecMul},$ we will also assume the existence of $\text{SecBeaGen},$ the secure protocol to generate the required triples. For simplicity of notation sometimes we write $\text{SecAdd}([u],[v])$ and $\text{SecMul}([u],[v])$ as $[u]+[v]$ and $[u]\cdot [v]$ respectively. We note that here we also assume the existence of $\text{SecMul3},$ which is a variant of $\text{SecMul}$ with three secretly shared inputs that outputs the secret share of the product of the three inputs.
    
    We further assume the existence of a protocol $\text{ZeroGen},$ which securely generates a random share of zero. Note that this simply means that it generates a uniformly random $x\in \mathbb{F}_q$ where we let $\mathcal{S}_1$ and $\mathcal{S}_2$ hold $x$ and $-x$ respectively. Such functionality is used along with $\text{SecAdd}$ to refresh a secretly shared value, i.e., generating a new secret share of the same value by introducing fresh randomness to remove the correlation between the shares of two secretly shared values. Lastly, we also assume the existence of a secure select protocol $\text{SecSel}$ such that given three secretly shared values $[v_0],[v_1], $ and $[i]$ where $i\in \{0,1\},$ it returns a fresh share of $v_i.$ Note that this can be realized by calculating $[y]\leftarrow [v_0]+[i]\cdot([v_1]-[v_0]).$ 
    
    We note that the functionalities above also include the case where the required input and outputs are in the form of vectors or matrices of values.
    \item Next, we consider the secure protocol $\text{SecSP},$ which, given two secretly shared vectors of the same length $n, [\mathbf{u}]$ and $[\mathbf{v}]$ outputs a secret share of the standard inner product between them. It is easy to see that $\text{SecSP}$ can be seen as a special case of $\text{SecMul}$ with inputs being an $n\times 1$ matrix obtained by treating $\mathbf{u}$ as a row vector and a $1\times n$ matrix obtained by treating $\mathbf{v}$ as a column vector.
    \item We assume the existence of a secure protocol $\text{SecRanGenInv},$ which securely generates a secretly shared random invertible value. Here we note that it can either return a secretly shared non-zero finite field element $[v]$ or a secretly shared invertible square matrix $[M]$ of any size. Note that this can be achieved by generating two of the random values and sacrificing one of them to check the invertibility of the other, as briefly discussed in the following. First, we call $\text{SecRanGen}$ twice to obtain $[u]$ and $[v].$ Having this, we call $\text{SecMul}([u],[v])$ to obtain $[w]$ where $w=uv$ and call $w\leftarrow\text{SecRec}([w]).$ Note that if $w$ is non-zero, then $u$ must also be non-zero and hence can be used as the output of $\text{SecRanGenInv}.$ Otherwise, a new pair $([u],[v])$ is generated and the check is repeated. Note that the same idea can be used to design the secure protocol generating a secretly shared invertible square matrix $[R].$
    \item We note that since values considered here are encoding of real numbers, we also require the secure protocol to be compatible with the encoding function $\varphi_q.$ For instance, this implies that in $\text{SecMul},$ it contains a secure truncation protocol to preserve the precision level of the product.
    \item We assume the existence of $\text{SecGE},$ a secure protocol that, given two secretly shared values $[u]$ and $[v],$ returns $[b]$ where $b=1$ if $u\geq v$ and $b=0$ otherwise. Here the comparison is done by treating $u$ and $v$ as $\bar{u},\bar{v}\in \left\{-\frac{q-1}{2},\cdots, \frac{q-1}{2}\right\}\subseteq \mathbb{Z}$ such that $u$ and $v$ are the classes of integers that equal to $\bar{u}\pmod{q}$ and $\bar{v}\pmod{q}$ respectively. In our work, such protocol is done by first converting $[u]$ to $[\mathbf{u}],$ the shared binary representation of $u$ and comparison can then be represented as a function of the component-wise addition and multiplication modulo $2$ \cite{demmler2015aby}. Note that with the existence of $\text{SecSel}$ and $\text{SecGE},$ it is then straightforward to design secure protocols $\text{SecMax}$ and $\text{SecSrt},$ which, given a list of secretly shared values returns a fresh share of the largest value and a list containing the fresh shares of the same values but in a sorted manner. Here we assume that both protocols can also take a secure comparison protocol $\Pi$ as one of its inputs in the case that the comparison needs to be done using a different definition from the one we define above. In this work, we assume that $\text{SecSrt}$ is constructed based on the bitonic sort proposed in \cite{Bat68}.
    \item We assume the existence of $\text{SecDiv},$ a secure protocol that, given two secretly shared values $[u]$ and $[v]$ for some $u,v\in \mathbb{F}_q,$ returns $[w]$ where $w=\lfloor\frac{u}{v}\rfloor.$ Note that it can be realized, for example following the protocol proposed in \cite{DNT12} or through binary representation conversion as described in \cite{demmler2015aby}.
\end{enumerate}

A summary of 2PC functionalities are outlined in Table \ref{tab:2pcfunc}.
\begin{table}[htb]
\centering
\caption{2PC Functionalities.}
\label{tab:2pcfunc}
\begin{tabular}{c|p{2.3in}}
\toprule
Functionality & Description\\
\hline
$\text{SecShare}$ & Produce secret share of the input finite field element\\
$\text{SecRec}$ & Recover a secretly shared value from its shares\\
$\text{SecRanGen}$ & Generate a secretly shared uniformly random finite field element\\
$\text{SecAdd}$ & Calculate a secret share of the sum of two secretly shared inputs\\
$\text{SecMul}$ & Calculate a secret share of the product of two secretly shared inputs\\
$\text{SecMul3}$ & Calculate a secret share of the product of three secretly shared inputs\\
$\text{SecSP}$ & Calculate a secret share of the standard inner product of two secretly shared vectors of the same length\\
$\text{ZeroGen}$ & Randomly generate a secret share of zero\\
$\text{SecSel}$ & Obliviously obtain a fresh share of a selected secret shared value out of two secretly shared input\\
$\text{SecRanGenInv}$ & Generate a secretly shared uniformly random non-zero finite field element\\
$\text{SecGE}$ & Calculate a secretly shared bit indicating the comparison result between two secretly shared values\\
$\text{SecMax},\text{SecSrt}$ & Calculate a secretly shared maximum value and sorted list given a secretly shared list of inputs\\
$\text{SecDiv}$ & Calculate the secret share of the result of the division of a secretly shared input by another secretly shared input \\
\bottomrule
\end{tabular}
\end{table}

\subsection{Complexities of ABY-based 2PC Functionalities}\label{app:2pc-funccomp}
    In this section, we provide a summary of the complexities of the basic 2PC building blocks which are directly based on ABY. Here, the size parameters in Table \ref{tab:2pcfunccomp} refers to the size of the inputs and outputs of the functionalities. We further let $\kappa$ be the security parameter and $q$ be the size of the underlying finite field. We recall that ABY-based 2PC contains three possible secret sharing forms, Arithmetic, Boolean and Yao sharing. This implies the need of $6$ conversion functionalities used to convert value shared in one form to any other form. Here we denote such conversion functionalities by $\text{SecA2B},\text{SecB2A},\text{SecA2Y},\text{SecY2A},\text{SecB2Y},$ and $\text{SecY2B}.$ Due to the application considered in this work, apart from the conversion functionalities $\text{SecB2A},\text{SecY2A},\text{SecB2Y},$ and $\text{SecY2B}$ as well as the $\text{SecDiv}^{(Y)},$ which is a secure division protocol using Yao's garbled circuit, we assume that the inputs and outputs are secretly shared using Arithmetic secret sharing scheme. As discussed in \cite{PSSY21}, $\text{SecDiv}^{(Y)}$ is realized by the secure division algorithm implemented in EMP-Toolkit \cite{emp-toolkit}, which takes $O(1)$ round with total communication bit of $O(\kappa \log^2 q)$ bits.

\begin{table*}[htbp!]
\centering
\caption{Complexities of 2PC Functionalities.}
\label{tab:2pcfunccomp}
\begin{tabular}{c|c|c|c|c}
\toprule
Functionality & Size Parameter & Offline Memory Requirement & Rounds & Communication Complexity (bits)\\
\hline
$\text{SecA2B}$ &$1$& $O(\kappa \log q)$ & $2$ & $O(\kappa \log q)$\\
$\text{SecB2A}$ &$1$& $O(\kappa \log q + \log^2 q)$ & $1$ & $O(\log q)$\\
$\text{SecA2Y}$ &$1$& $O(\kappa \log q)$ & $1$ & $O(\kappa \log q)$\\
$\text{SecY2A}$ &$1$& $O(\kappa \log q)$ & $1$ & $O(\log q)$\\
$\text{SecB2Y}$ &$1$& $O(\kappa \log q)$ & $1$ & $O(\kappa \log q)$\\
$\text{SecY2B}$ &$1$& $O(\log q)$ & $2$ & $O(\log q)$\\
\hline
$\text{SecShare}$ & $m\times n$ & $0$ & $1$ & $O(mn\log q)$\\
$\text{SecRec}$ & $m\times n$ & $0$ & $1$ & $O(mn\log q)$\\
$\text{SecRanGen}$ &$m\times n$ & $0$ & $1$ & $O(mn\log q)$ \\
$\text{SecAdd}$ &  $m\times n$ & $0$ & $0$ & $0$\\
$\text{SecMul}$ & $m\times n, n\times p$ & $O(mp\log q)$ & $1$ & $O(mp\log q)$\\
\multirow{2}{*}{$\text{SecMul3}$} & $1$ & $O(\log q)$ & $1$ & $O(\log q)$  \\
& $m\times n, n\times p, p\times r$ &$O((mp+mr)\log q)$ & $2$ & $O(mp\log q)$ \\
$\text{SecMul4}$ & $1$ & $O(\log q)$ & $1$ & $O(\log q)$\\
$\text{SecSP}$ & $n$ & $O(\log q)$ & $1$ & $O(\log q)$\\
$\text{ZeroGen}$ & $m\times n$ & $0$ & $1$ & $O(mn\log q)$\\
$\text{SecGE}$ & 1 & $O(\kappa\log q \log \log q )$&$O(\log q)$ & $O(\log^2q \log\log q )$ \\
$\text{SecDiv}^{(Y)}$&$1$&$0$&$O(1)$&$O(\kappa \log^2 q)$\\
\bottomrule
\end{tabular}
\end{table*}

\subsection{Complexities of Functionalities based on Basic ABY-based 2PC} \label{app:2pc-funccomb}
    In this section, we provide a summary of several functionalities that we may construct based on the basic 2PC building blocks previously discussed. We note that, as previously discussed, since $\text{SecRanGenInv}$ repeats the pair generation phase and verification phase if the verification fails, the number of building blocks discussed in Table \ref{tab:2pcfunccomb} is based on its expected value. Since the probability of having both generated numbers to be zero is $\left(\frac{q-1}{q}\right)^2,$ in expectation, the total number of iterations is $2.$ 
    \begin{table*}[htbp!]
\centering
\caption{Intermediate Functionalities}
\label{tab:2pcfunccomb}
\begin{tabular}{p{1in}|c|p{1in}| p{2in}}
\toprule
Functionality & Size Parameter & Additional Offline Memory Requirement & Building Blocks\\
\hline
$\text{SecRanGenInv}$ & $m\times m$ & $0$  & $4$ calls of $\text{SecRanGen} (m\times m), 2$ calls of $\text{SecMult} (m\times m), 2$ calls of $\text{SecRec}$ \\
\hline
$\text{SecMI}$ & $m\times m$ &$O(m^2\log q)$& $2$ calls of $\text{SecMul} (m\times m),1$ call of $\text{SecRec}$\\
\hline
$\text{SecSel}$ & $m\times n$ & $0$ & $1$ call of $\text{SecMul}$ \\
\hline
$\text{SecSrt}$ based on comparison function $\Pi$ & $T$ & $0$ & $O\left(Tlog^2(T)\right)$ calls of $\Pi$ and $O\left(T\log^2(T)\right)$ calls of $\text{SecSel}$ \\
\hline
$\text{SecMax}$ based on comparison function $\Pi$ & $T$ & $0$ & $O(\log T)$ parallel calls of $\Pi$ and $\text{SecSel}$ totalling $O(T)$ calls of $\Pi$ and $\text{SecSel}$\\
\hline
$\text{SecDiv}$ & $1$ & $0$ & $1$ call to $\text{SecA2Y}, 1$ call to Yao's based division algorithm  and one call to $\text{SecY2A}$\\
\bottomrule
\end{tabular}
\end{table*}

\section{Secure Round Selection without Client Pre-computation}\label{app:SRS}
In this section, we discuss $\text{SecRS}^{(alt)},$ a variant of $\text{SecRS}$ where $\|g_i^t\|$ is not pre-computed and provided by the client. Note that this variant takes advantages of the fact that $\|g_i^t\|^2 = \langle g_i^t,g_i^t\rangle.$ Furthermore, we also note that given two values $a$ and $b,$ we have $a\geq b$ if one of the following is satisfied:
\begin{itemize}
    \item $a>0$ and $b<0$ 
    \item If $a$ and $b$ have the same sign, let $w=1$ if $a>0$ and $w=0$ otherwise. Then $a\geq b$ if $(2w-1) \cdot (2\cdot (a^2\geq b^2)-1) = 1$ where $(a\geq b)=1 $ if $a\geq b$ and $0$ otherwise.
\end{itemize}

So in order to compare two pairs $A_i=(\mathbf{x}_i,\mathbf{y}_i)$ and $A_j= (\mathbf{x}_j,\mathbf{y}_j)$ where $A_i\succcurlyeq A_j$ if and only if $\frac{\langle \mathbf{x}_i,\mathbf{y}_i\rangle}{\|\mathbf{x}_i\|\cdot \|\mathbf{y}_i\|},$ we may follow the following strategy:
\begin{itemize}
    \item Calculate $u_i = \langle \mathbf{x}_i,\mathbf{y}_i\rangle, v_i=\langle \mathbf{x}_i,\mathbf{x}_i\rangle \cdot \langle \mathbf{y}_i,\mathbf{y}_i\rangle, w_i=(u_i>0), a_i = u_i^2$ and store $B_i = (w_i, a_i,v_i).$ Here $(u_i>0)$ returns $1$ if $u_i\geq 0$ and $0$ otherwise. Perform similar calculation to $A_j$ to obtain $B_j=(w_j,a_j,v_j).$ 
    \item Then $A_i\succcurlyeq A_j$ if and only if $w_i>w_j$ or $2(w_i-1)\cdot \left(2\left(\frac{a_i}{v_i}>\frac{a_j}{v_j}\right)-1\right)=1.$
\end{itemize}

We further define $\text{SecGE3}$ and $\text{SecGE4}$ in order to help in the comparison between $B_i$ and $B_j.$ Note that here, similar to the discussion of $\text{SecRS},$ the algorithm $\text{SecGE3}$ is used if we first calculate $b_i\triangleq \frac{a_i}{v_i}$ and comparison is done between $C_i\triangleq (w_i,b_i)$ and $C_j\triangleq (w_j,b_j).$ On the other hand $\text{SecGE4}$ is used when we try to avoid the division operation and comparison is done on $B_i$ and $B_j$ directly.

\begin{algorithm}
\SetAlgoNoEnd
\caption{$\text{SecGE3}$}
\label{alg:secge3}
\KwIn{$([w_1],[b_1]), ([w_2],[b_2])$}
\KwOut{$[b]$ where $b\in\{0,1\}$ and $b=1$ if $w_1\sqrt{b_1}\geq w_2\sqrt{b_2}.$}
Calculate $[x]\leftarrow \text{SecGE}([w_i],[w_j])$ and $[x']\leftarrow \text{SecGE}([w_j],[w_i])$\;
Calculate $[z]\leftarrow \text{SecGE}([b_i],[b_j]), [z']\leftarrow \text{SecAdd}(2\text{SecMult}([w_i],[z_i]),-\text{SecAdd}([z],[w_i])$\;
Calculate $[b]\leftarrow \text{SecAdd}([x],\text{SecMult3}([x],[x'],[z']))$\;

\KwRet{$[b]$}
\end{algorithm}
It is then easy to see that $\text{SecGE4}$ can be easily constructed from Algorithm \ref{alg:secge3} by replacing the calculation of $[z]$ by $[z]\leftarrow \text{SecGE2}(([a_i],[v_i]),([a_j],[v_j])).$

By the discussion above, we propose $\text{SecRS}^{(alt)},$ as can be observed in Algorithm \ref{alg:round_selection_alt}.

It can then be verified that compared to $\text{SecRS}$ described in Algorithm \ref{alg:round_selection}, $\text{SecRS}^{(alt)}$ has the following additional calculations. We divide the protocol into two phases, namely the construction of the $\mathcal{L}$ phase and the sorting phase. It is then easy to verify that the construction phase incurs an additional $1$ more call each of $\text{SecSP},\text{SecGE},\text{SecMul}$ for the construction of $\mathcal{L}.$ Furthermore, for the sorting phase, replacing $\text{SecGE2}$ by $\text{SecGE4}$ or $\text{SecGE}$ by $\text{SecGE3}$ incurs additional two calls of $\text{SecGE}$ and $1$ call each of $\text{SecMult}$ and $\text{SecMult3}.$ As previously discussed, since bitonic sort \cite{Bat68} is used, it requires $\frac{\ell_T^2+\ell_T}{4}T$ comparison to be done. So the sorting phase incurs an additional of $\frac{\ell_T^2+\ell_T}{2}T$ calls of $\text{SecGE}$ and $\frac{\ell_T^2+\ell_T}{4}T$ calls each of $\text{SecMult}$ and $\text{SecMult3}.$

\begin{algorithm}[h!]
\SetAlgoNoEnd
\caption{Alternate Secure Round Selection ($\text{SecRS}^{(alt)}$)}
\label{alg:round_selection_alt}
\KwIn{Secretly shared historical gradients $\{[G_i]\}^T,$ historical global models $\{M_i\}^T,$ the initial global model $M_0,$ the number of historical rounds $T,$ the number of selected rounds $T^{\prime},$ the switching threshold $T_{\delta},$ the target client $c^t.$}
\KwOut{Selected historical gradients $\{[G_i]\}^{T^{\prime}}$ along with corresponding selected historical global models $\{M_i\}^{T^{\prime}}$.}
Obtain $\{[g_i^t]\}^T$ from $\{[G_i]\}^T$\;
$[\mathcal{L}]\leftarrow \epsilon$\; \tcp{Initialize $\mathcal{L}$ as an empty list} 
\eIf{$T \leq T_{\delta}$}{
\For{$i \leftarrow 1$ \KwTo $T$ }{
$[u_i]\leftarrow \text{SecSP}([g_i^t],M_i-M_{i-1})$ \;
$[v_i]\leftarrow \text{SecMul}(\text{SecSP}([g_i^t],[g_i^t]),\|M_i-M_{i-1}\|^2)$ \;
$[w_i]\leftarrow \text{SecGE}([u_i],0)$ and $[a_i]\leftarrow \text{SecMul}([u_i],[u_i]);$
$[\mathcal{L}]\leftarrow [\mathcal{L}]\|(i,[w_i],[a_i],[v_i])$ \tcp{Append $(i,[w_i],[a_i],[v_i])$ to $[\mathcal{L}]$}
}
$[\mathcal{L}']\leftarrow \text{SecSrt}([\mathcal{L}],\text{SecGE4})$\; \tcp{Sort $\mathcal{L}$ based on the ``$\succcurlyeq$'' rule defined in $\text{SecGE4}$ and store it in $[\mathcal{L}']$}
}
{
\For{$i \leftarrow 1$ \KwTo $T$ }{
$[u_i]\leftarrow \text{SecSP}([g_i^t],M_i-M_{i-1})$ \;
$[v_i]\leftarrow \text{SecMul}(\text{SecSP}([g_i^t],[g_i^t]),\|M_i-M_{i-1}\|^2)$ \;
$[w_i]\leftarrow \text{SecGE}([u_i],0)$ and $[a_i]\leftarrow \text{SecMul}([u_i],[u_i]);$
$[b_i]\leftarrow \text{SecDiv}([a_i],[v_i]);$
$[\mathcal{L}]\leftarrow [\mathcal{L}]\|(i,[w_i],[b_i])$
}
$[\mathcal{L}'] \leftarrow \text{SecSrt}([\mathcal{L}],\text{SecGE3})$\;
}
$\mathbf{d}\leftarrow \epsilon$\;
\For{$i\leftarrow 1$ \KwTo $T'$}
{$\mathbf{d}\leftarrow \mathbf{d}\|\text{SecRec}([\mathcal{L}'[i,0]])$\; \tcp{The index of the $i$-th largest value is recovered from the first element of the $i$-th entry of the sorted list $\mathcal{L}'$}}
Obtain $\{[G_i]\}^{T^{\prime}}, \{M_i\}^{T^{\prime}}$ from $\{[G_i]\}^T$ based on $\mathbf{d}$\;
\KwRet{$\{[G_i]\}^{T^{\prime}}, \{M_i\}^{T^{\prime}}$}
\end{algorithm}

\section{Round and Invocation Complexities of the Starfish Scheme}\label{app:RouInComp}
Now we consider the round and invocation complexities of our proposed schemes.
\begin{itemize}
    \item Threshold determination: Note that in order to facilitate the unlearning step, for each client $c^j, \delta^j$ is defined to be the maximum among $\delta_1^j,\cdots, \delta_T^j.$ This is done by calling $\text{SecMax}$ with input $[\delta_1^j],\cdots, [\delta_T^j].$ Hence it requires $n$ calls of $\text{SecMax}$ with input size $T$ done in parallel. 
    \item Secure Round Selection ($\text{SecRS}$): Here we note that $\text{SecRS}$ has two possible cases depending on whether $T<T_\delta.$ If Method $1$ is used, lines $4$ up to $8$ of Algorithm \ref{alg:round_selection} will be used where it requires $T$ calls of $\text{SecSP}$ with input of length $m$ and $T$  calls of $\text{SecMul}$ each done in parallel. Furthermore, it requires $1$ call of $\text{SecSrt}$ with the underlying comparison functionality $\text{SecGE2}.$ Assuming that Bitonic Sort \cite{Bat68} is used, it requires $O(\log^2 T)$ parallel calls of $\text{SecGE2}$ and $\text{SecSel}$ totalling to $O(T\log^2 T)$ calls of $\text{SecGE2}$ and $O(T\log^2 T)$ calls of $\text{SecSel}.$ Note that $\text{SecGE2}$ consists of $2$ calls of $\text{SecMul}$ done in parallel and $1$ call of $\text{SecGE}.$ Note that $\text{SecSP},\text{SecMul}$ and $\text{SecSel}$ require one round of communication each while $\text{SecGE}$ requires $O(\log^2 T)$ rounds of communication. Then we can have the following conclusion about lines $4$ up to $8$ of Algorithm \ref{alg:round_selection}. The total number of communication round required contains $O(\log^2 T)$ rounds of $\text{SecGE}, 1$ round of $\text{SecSP}, O(\log^2 T)$ rounds of $\text{SecMul}$ and $O(\log^2 T)$ rounds of $\text{SecSel}.$ The total invocation complexities for lines $4$ up to $8$ of Algorithm \ref{alg:round_selection} is $T$ calls of $\text{SecSP}, O(T\log^2 T)$ calls of $\text{SecMul}, O(T\log^2 T)$ calls of $\text{SecSel}$ and $O(T\log^2 T)$ calls of $\text{SecGE}.$

    On the other hand, if Method $2$ is used, lines $10$ up to $15$ of Algorithm \ref{alg:round_selection} will be considered instead. It is easy to see that it requires $T$ calls of $\text{SecSP}$ with input of length $m, T$  calls of $\text{SecMul}$ each done in parallel, $T$ calls of $\text{SecDiv}$ each done in parallel and $1$ call of $\text{SecSrt}$ with the underlying comparison functionality $\text{SecGE}.$ As before, with Bitonic sort being used as the secure sorting algorithm, we have the following conclusion regarding lines $10$ up to $15$ of Algorithm \ref{alg:round_selection}. The total number of communication round required contains $O(\log^2 T)$ rounds of $\text{SecGE}, 1$ round of $\text{SecSP}, 1$ round of $\text{SecMul}, 1$ round of $\text{SecDiv},$ and $O(\log^2 T)$ rounds of $\text{SecSel}.$ The total invocation complexities for lines $10$ up to $15$ of Algorithm \ref{alg:round_selection} is $T$ calls of $\text{SecSP}, T$ calls of $\text{SecMul}, T$ calls of $\text{SecDiv}, O(T\log^2 T)$ calls of $\text{SecSel}$ and $O(T\log^2 T)$ calls of $\text{SecGE}.$
    Lastly, lines $16$ up to $20$ require $1$ round of communication where $T'$ calls of Algorithm \ref{alg:round_selection} are done in parallel.
    \item Secure Update Estimation ($\text{SecUE}$): We note that for line $2$ of $\text{SecUE},$ which can be found in Algorithm \ref{alg:update_estimation}, the clients performs $1$ round of communication where the shares of $\Delta M_{t_i}$ and $\Delta G_{t_i}$ are sent to the servers. The computations are then done by the servers. Here we note that each iteration of the outer loop must be done sequentially. On the other hand, it is easy to verify that for each iteration, which consists of lines $4$ up to $13$ of Algorithm \ref{alg:update_estimation}, requires $n-1$ calls of $\text{SecMI}$ done in a parallel, $3(n-1)$ calls of $\text{SecMul3}$ divided to $2$ rounds. Next, it requires $n-1$ calls of $\text{SecMul}$ in a parallel. The iteration ends by having $1$ call of $\text{SecRec}.$ So in conclusion, the total number of rounds is $B$ rounds of $\text{SecMI}, 2B$ rounds of $\text{SecMul3}, B$ rounds of $\text{SecMul}$ and $B$ rounds of $\text{SecRec}.$ On the other hand, its total number of invocations are $B(n-1)$ calls of $\text{SecMI},3B(n-1)$ calls of $\text{SecMul3}, B(n-1)$ calls of $\text{SecMul}$ and $B$ calls of $\text{SecRec}.$

    \item Secure Threshold Checking ($\text{SecTC}$): It is easy to see that $\text{SecTC},$ which can be found in Algorithm \ref{alg:threshold_checking}, requires $(n-1)m$ calls of $\text{SecSP}, (n-1)m$ calls of $\text{SecMult},(n-1)m+n$ calls of $\text{SecGE},$ and one call of $\text{SecRec}$ where the calls of $\text{SecSP}$ and $\text{secMult}$ are done in parallel in one invocation round, the $\text{SecGE}$ calls are divided to two invocation rounds. Now, noting that the communication round of $\text{SecSP}$ is at least that of $\text{SecMult}$ (in fact with the use of ABY, the two functionalities have the same number of communication rounds, namely $1$), in total, $\text{SecTC}$ takes $1$ invocation round of $\text{SecSP}, 2$ invocation rounds of $\text{SecGE}$ and one invocation rounds of $\text{SecRec}$ totaling to $O(nm)$ invocations of $\text{SecSP}, O(nm)$ invocations of $\text{SecMult}, O(nm)$ invocations of $\text{SecGE}$ and $1$ call of $\text{SecRec}.$
\end{itemize}
Having the main sub-functionalities, we are now ready to discuss the complexity of the proposed the Starfish scheme discussed in Algorithm \ref{alg:starfish}. Here we divide the analysis to two stages; The first stage is the learning stage, which is lines $1$ up to $9,$ which consists of $T$ rounds of the Federated Learning scheme. The second stage starts when an unlearning request is received. Here it contains lines $10$ onwards where the threshold is determined, $T'$ rounds are selected, and $T'$ rounds of unlearning are executed. 
\begin{itemize}
    \item Learning stage: It is easy to see that the $T$ learning iterations must be done sequentially while each iteration requires each client to call $3$ invocations of $\text{SecShare}$ in parallel and the servers to call $\text{SecRec}.$ Hence in total, the learning stage requires $T$ rounds of $\text{SecShare}$ and $T$ rounds of $\text{SecRec}$ totalling to $3T$ invocations of $\text{SecShare}$ for each client and $T$ invocations of $\text{SecRec}$ by the servers.
    \item Unlearning stage: Suppose that $c^t$ sends an unlearning request. Here we consider the complexity of the unlearning stage in the worst case, which happens when for any $i=1,\cdots, T', t_i\equiv 0\pmod{T_c}$ and $f_{t_i}^j =1$ for any $j\neq t.$ In such case, it is easy to see that the unlearning stage requires $n$ calls of $\text{SecMax}$ with input of size $T$ by the servers which are done in parallel, $1$ call of $\text{SecRS},$ and $T'$ calls each of $\text{SecUE}$ and $\text{SecTC}$ by the servers and $\text{SecShare}$ by each client other than the requester $c^t$ which are all done in sequential manner. So in total, the unlearning stage requires $1$ round of $\text{SecMax}, 1$ round of $\text{SecRS}, T'$ rounds of $\text{SecUE},$ and $T'$ rounds of $\text{SecTC}$ by the servers and $T'$ round of $\text{SecShare}$ done by each client other than the requester totaling of up to $n$ invocations of $\text{SecMax}, 1$ invocation of $\text{SecRS}, T'$ invocations of $\text{SecUE},$ and $T'$ invocations of $\text{SecTC}$ by the servers while we have in total $T'$ invocations of $\text{SecShare}$ by clients other than the requester.
\end{itemize}

\section{Secure Matrix Inversion for Secure Update Estimation}\label{app:SecMI}
In this section, we briefly discuss $\text{SecMI},$ the protocol used in Algorithm \ref{alg:update_estimation} to inverse a secretly shared square invertible matrix. 
SecMI introduces operations in the offline phase for generating 2PC materials. More specifically, it requires the generation of a secure invertible square matrix $[R]$ with the same dimension as $X$ through $\text{SecRanGenInv}$ as well as the values required in the call of $\text{SecMul}$ done in Step $1.$ We note that the call of $\text{SecMul}$ in step $3$ does not require any preprocessed values nor communication.
\begin{algorithm}
\SetAlgoNoEnd
\caption{Secure Matrix Inversion (SecMI)}
\label{alg:matrix_inversion}
\KwIn{An invertible square matrix $[X],$ a randomly generated invertible square matrix $[R]$ of the same dimension as the matrix $X$ through \text{SecRandGenInv}.}
\KwOut{The inverse of the input matrix $[X^{-1}]$.}
$N= \text{SecRec}(\text{SecMul}([X],[R]))$\;
Calculate $N^{-1}$ from $N$ in the clear\;
$[X^{-1}]=\text{SecMul}([R],N)$\;
\KwRet{$[X^{-1}]$}
\end{algorithm}

It is easy to see that Algorithm \ref{alg:matrix_inversion} correctly calculates $[X^{-1}]$ given $[X].$ Furthermore, since $R$ is uniformly sampled from the space of any invertible matrices, the distribution of $N=XR$ is independent of the value of $X,$ hence the reveal of $N$ does not reveal any information about $X,$ proving its security. 
\section{Difference in Unlearned Models between Starfish and Train-from-Scratch}
\label{app:ModDif}

In this section, we first outline the assumptions that support our theoretical analysis. Subsequently, we demonstrate that the difference between the unlearned global model obtained through the Starfish approach and that achieved via the train-from-scratch method can be bounded.


\begin{assumption}[Strong convexity and smoothness]\label{ass_smooth}
    The function $F$ is $\mu$-strongly convex and $L$-smooth for a positive coefficient $\mu$, i.e., $\forall M_1, M_2\in \mathbb{R}^d$, the following equation holds:
    \begin{equation}
        F(M_1) \leq F(M_2) + \nabla F(M_2)^\top (M_1 - M_2) + \frac{L}{2} \Vert M_1 - M_2 \Vert^2,
    \end{equation}
    \begin{equation}
        F(M_1) \geq F(M_2) + \nabla F(M_2)^\top (M_1 - M_2) + \frac{\mu}{2} \Vert M_1 - M_2 \Vert^2.
    \end{equation}
\end{assumption}
\begin{assumption}\label{ass_hessian}
    Function $F: \mathbb{R}^d \rightarrow \mathbb{R}$  is twice continuously differentiable, $L$-smooth, and $\mu$-strongly convex, i.e.,
    \begin{equation}
        \mu I \leq \nabla^2 f(M) \leq LI.
    \end{equation}
    where $I \in \mathbb{R}^d $ and $\nabla^2 f(M)$ is the Hessian of gradient.
\end{assumption}

In the Starfish scheme, we employ a round selection strategy with a selection rate of $\sigma$. Consequently, it is essential to show that the model obtained through Starfish at the time $t_i$ and the one achieved by the train-from-scratch method at $t=\lceil \frac{1}{\sigma}t_i \rceil$ exhibit a bounded difference.

Recall that the global model obtained is updated as follows:

\begin{equation}
    \begin{aligned}
    \hat{M}_{t_i+1} &= \hat{M}_{t_{i}} -\eta_u \hat{G}_{t_{i}} \\
    &= \hat{M}_{t_{i}} -\eta_u \mathcal{H}_{t_{i}}^{-1} G_{t_{i}}.
\end{aligned}
\end{equation}
Since $f(\hat{M})$ is $\mu$-strongly convex, according to the Assumption \ref{ass_smooth}, we can have:
\begin{equation}\label{eq_convex}
    F(\hat{M}_{t_i+1}) \leq F(\hat{M}_{t_i}) - \eta_u \nabla_{t_i}^\top \triangle_{t_i} + \frac{\eta_u^2 L \Vert \triangle_{t_i} \Vert^2}{2},
\end{equation}
where $\triangle_{t_i} = \mathcal{H}_{t_{i}}^{-1} \nabla_{t_i} $ and $\nabla_{t_i} = \nabla F(\hat{M}_{t_i})$. For simplicity, let $\omega_t=\sqrt{\nabla_{t_i}^\top \mathcal{H}_{t_{i}}^{-1} \nabla_{t_i}}$. Thus, $\omega_t^2=\triangle_{t_i}^\top \mathcal{H}_{t_{i}}^{\top} \triangle_{t_i}$. According to the Assumption \ref{ass_hessian}, we can know that $\omega_t^2 \geq \mu \Vert \triangle_{t_i} \Vert^2$. Thus, Eq. \ref{eq_convex} can be computed as follows:
\begin{equation}\label{eq_convex1}
     F(\hat{M}_{t_i+1}) \leq F(\hat{M}_{t_i}) - \eta_u \omega_t^2 + \frac{L}{2\mu} \eta_u^2 \omega_t^2.
\end{equation}
Then, we assume that unlearning rate $\eta_u = \frac{\mu}{L} = \eta$. Therefore, Eq. \ref{eq_convex1} can be:
\begin{equation}
    F(\hat{M}_{t_i+1}) \leq F(\hat{M}_{t_i}) - \frac{1}{2} \eta \omega_t^2.
\end{equation}
Thus, we rearrange the term: 
\begin{equation}
    \eta \omega_t^2 \leq 2 [F(\hat{M}_{t_i})-F(\hat{M}_{t_i+1}) ].
\end{equation}
Since $\omega_t^2 \geq \mu \Vert \triangle_{t_i} \Vert^2$, we can get:
\begin{equation}
    \eta \Vert \triangle_{t_i} \Vert^2 \leq \frac{2}{\mu} [F(\hat{M}_{t_i})-F(\hat{M}_{t_i+1}) ].
\end{equation}
When to iterate $t_i$ rounds, we have
\begin{equation}
\begin{aligned}
     \sum^{t_i-1}_{t_i=0} \eta \Vert \triangle_{t_i} \Vert^2 &\leq \frac{2}{\mu} [F(\hat{M}_{0})- F(\hat{M}_{t_i-1})] \\
     &\leq \frac{2}{\mu} [F(\hat{M}_{0})- F({M}^*)].
\end{aligned}
\end{equation}
where $M^*$ is the optimal solution for the function $F(M)$. Then, by applying the Cauchy-Schwarz inequality, we can get
\begin{equation}
\begin{aligned}
        \Vert \hat M_{t_i} -  M_0 \Vert &\leq \eta \sum_{t_i=0}^{t_i-1} \Vert \hat{G}_{t_i}\Vert \\
        & \leq \sqrt{\sum_{t_i=0}^{t_i-1} \eta \sum_{t_i=0}^{t_i-1} \eta \Vert \triangle_{t_i} \Vert^2 } \\
        &\leq \sqrt{\sum_{t_i=0}^{t_i-1} \frac{2 \eta}{\mu} [F(\hat{M}_{0})- F({M}^*)}].
\end{aligned}
\end{equation}

In train-from-scratch, the global model obtained is updated as follows:
\begin{equation}
    {M}_{t+1} = {M}_t -\eta_u G_{t}.
\end{equation}
According to Assumption \ref{ass_smooth}, we can have
\begin{equation}\label{eq_convex_retrain}
    F({M}_{t+1}) \leq F({M}_{t}) - \eta_u \nabla F(M_{t})^\top \nabla F(M_{t})+ \frac{\eta_u^2 L \Vert \nabla F(M_{t}) \Vert^2}{2} .
\end{equation}
We assume that unlearning rate for train-from-scratch is the same as that used in Starfish, which is $\eta=\frac{\mu}{L}$. Then, we can get:
\begin{equation}
    F({M}_{t+1}) \leq F({M}_{t}) - (\frac{\mu}{2}-1) \eta \Vert \nabla F(M_{t}) \Vert^2.
\end{equation}
Thus, by rearranging the term, we can have:
\begin{equation}
    \Vert \eta \nabla F(M_{t}) \Vert^2 \leq \frac{2}{\mu-2} [F({M}_{t})-F({M}_{t+1})].
\end{equation}
When to iterate $t$ rounds, we can obtain:
\begin{equation}
    \begin{aligned}
        \eta \sum_{t=0}^{t-1} \Vert \nabla F(M_{t}) \Vert^2 \leq \frac{2}{\mu-2} [F({M}_{0})-F({M^*})].
    \end{aligned}
\end{equation}
Similarly, by applying the Cauchy-Schwarz inequality, we can compute
\begin{equation}
    \begin{aligned}
        \Vert M_t-M_0 \Vert &\leq \eta \sum_{t=0}^{t-1} \Vert G_t \Vert \\
        &\leq \sqrt{\sum_{t=0}^{t-1} \eta \sum_{t=0}^{t-1} \eta \Vert \nabla F(M_{t}) \Vert^2} \\
        &\leq \sqrt{\sum_{t=0}^{t-1} \frac{2 \eta}{\mu-2} [F({M}_{0})-F({M^*})]}.
    \end{aligned}
\end{equation}
Finally, the difference between $\hat{M}_{t_i}-M_t$ can be:
\begin{equation}
    \begin{aligned}
        \Vert \hat{M}_{t_i}-M_t \Vert &\leq \Vert \hat{M}_{t_i} - \hat M_0\Vert + \Vert M_t - M_0\Vert \\
        &\leq 2\sqrt{ \eta [\frac{1}{\mu} + \frac{1}{\sigma(\mu-2)}][F({M}_{0})-F({M^*})] t_i  }.
    \end{aligned}
\end{equation}
where $\hat{M}_0 = {M}_0$ since the recovery process begins with the same global model. The above inequality shows that there exists an upper bound between the global model obtained by Starfish and that obtained by train-from-scratch.

\section{Additional Experimental Results}
\label{sec:add_exp}

This section includes additional experimental results, encompassing the unlearning performance across various Starfish parameter configurations, as well as an evaluation of 2PC efficiency under different Starfish parameters across diverse datasets.

\subsection{Additional experiments on unlearning performance}

\begin{figure}[htbp!]
    \centering
    \begin{subfigure}{0.48\linewidth}
        \centering
        \includegraphics[width=1\linewidth]{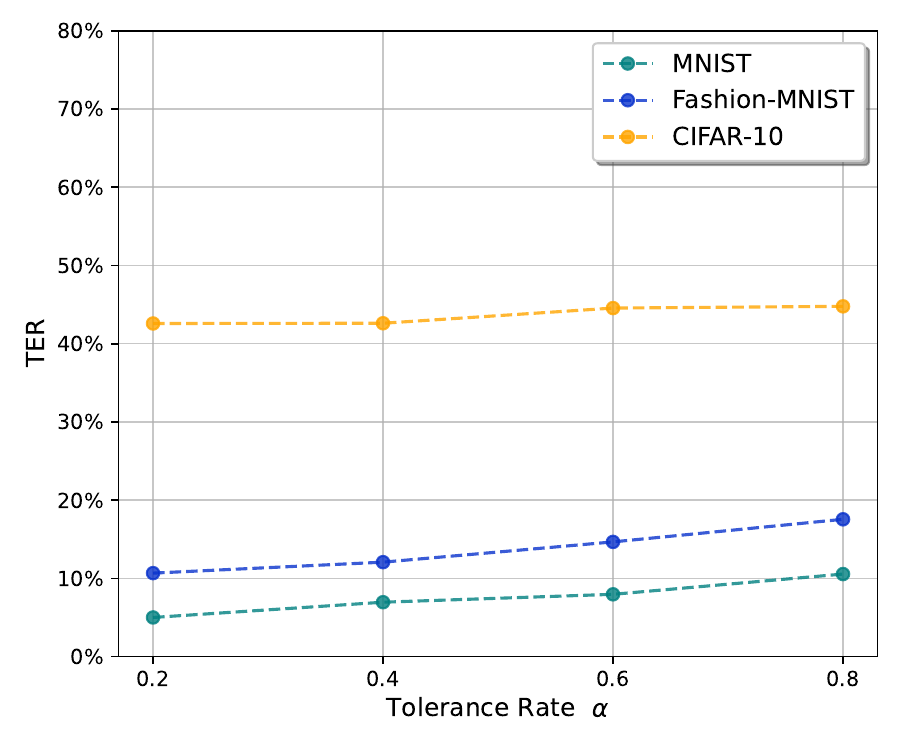}
        \caption{TER}
        \label{fig:tor_rt_ter}
    \end{subfigure}
    \begin{subfigure}{0.48\linewidth}
        \centering
        \includegraphics[width=1\linewidth]{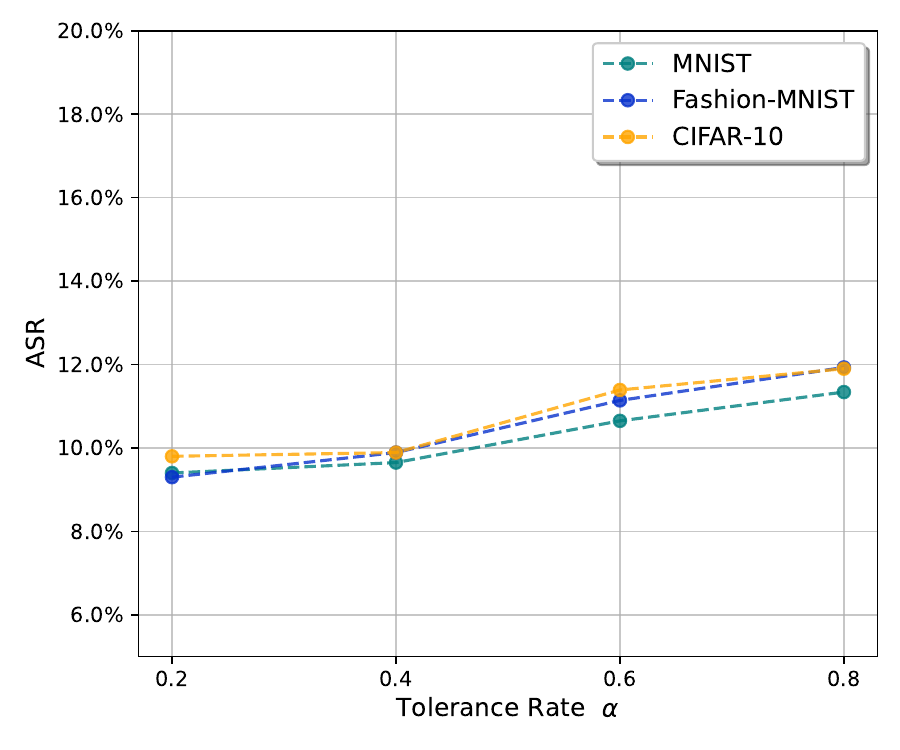}
        \caption{ASR}
        \label{fig:tor_rt_asr}
    \end{subfigure}

    \caption{Impact of the tolerance rate $\alpha$ on the unlearning performance of the Starfish scheme.}
    \label{fig:tor_rt_impact}
\end{figure}

\begin{figure}[htbp!]
    \centering
    \begin{subfigure}{0.48\linewidth}
        \centering
        \includegraphics[width=1\linewidth]{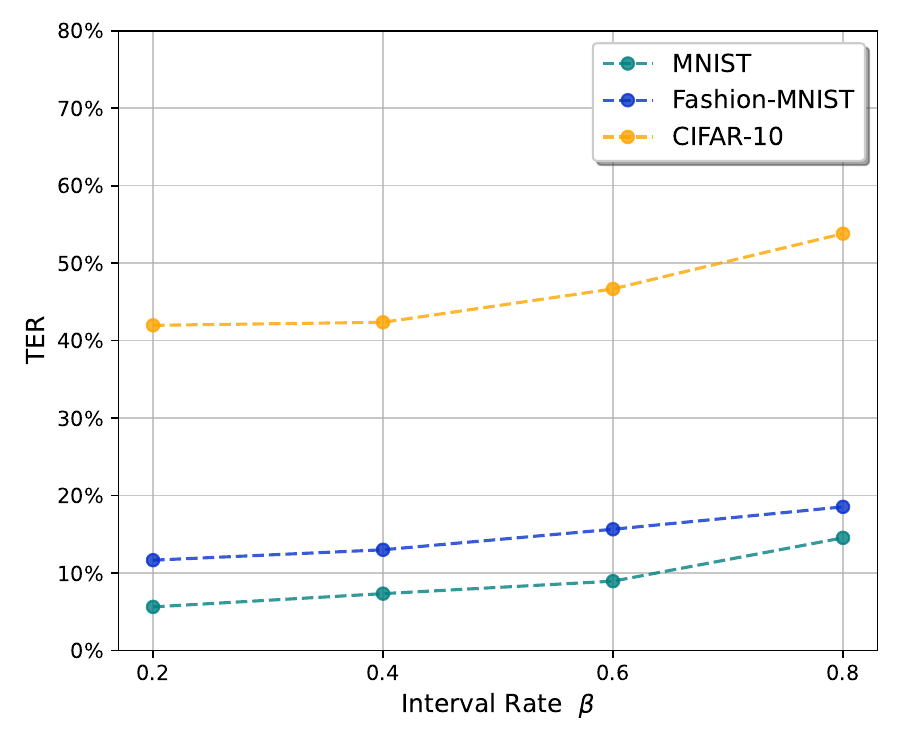}
        \caption{TER}
        \label{fig:int_rt_mia}
    \end{subfigure}
    \begin{subfigure}{0.48\linewidth}
        \centering
        \includegraphics[width=1\linewidth]{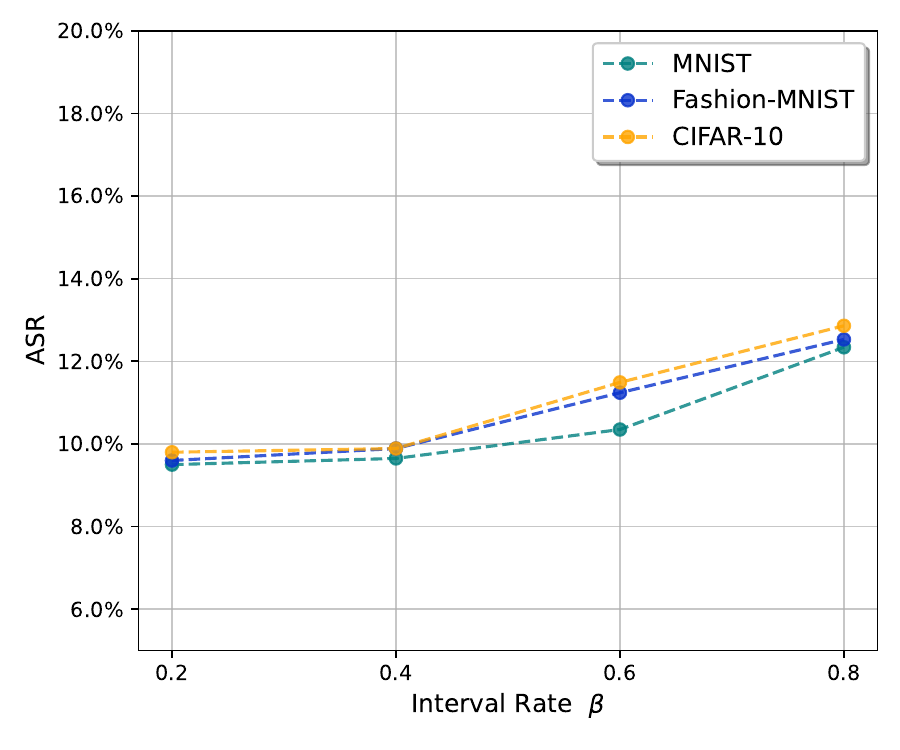}
        \caption{ASR}
        \label{fig:int_rt_ba}
    \end{subfigure}

    \caption{Impact of the interval rate $\beta$ on the unlearning performance of the Starfish scheme.}
\end{figure}

\subsection{Additional experiments on 2PC performance}

\begin{table}[htb]
\small
\caption{Comparison of 2PC performance for each step in a single unlearning round on Fashion-MNIST and CIFAR-10 dataset.}
\label{tab:step_one_round_FC}
\centering
\begin{tabular}{llcccc}
\hline \\[-1em]
\multicolumn{1}{c}{Step} & \multicolumn{1}{c}{Dataset} & \multicolumn{1}{c}{\begin{tabular}[c]{@{}c@{}}Offline\\ Time {\tiny (s)}\end{tabular}} & \multicolumn{1}{c}{\begin{tabular}[c]{@{}c@{}}Online\\ Time {\tiny (s)}\end{tabular}} & \multicolumn{1}{c}{\begin{tabular}[c]{@{}c@{}}Offline\\ Comm {\tiny (GB)}\end{tabular}} & \multicolumn{1}{c}{\begin{tabular}[c]{@{}c@{}}Online\\ Comm  {\tiny (GB)}\end{tabular}} \\ \hline \\[-1em]
$\text{TD}^1$ & {\scriptsize F-MNIST}  & 0.05 & 0.04 & 4.20 & 0.12 \\
    & {\scriptsize CIFAR-10}           & 0.06 & 0.04 & 4.20 & 0.12 \\
$\text{TD}^2$ & {\scriptsize F-MNIST}  & 0.01 & 0.03 & 0.11 & 0.01 \\
    & {\scriptsize CIFAR-10}           & 0.01 & 0.03 & 0.11 & 0.01 \\
$\text{RS}^1$ & {\scriptsize F-MNIST}  & 174.01 & 31.39 & 35.08 & 0.66 \\
    & {\scriptsize CIFAR-10}           & 274.51 & 57.77 & 51.58 & 0.97 \\
$\text{RS}^2$ & {\scriptsize F-MNIST}  & 104.48 & 18.90 & 21.05 & 0.39  \\
    & {\scriptsize CIFAR-10}           & 164.79 & 34.73 & 30.95 & 0.58 \\
$\text{UE}$  & {\scriptsize F-MNIST}   & 3.73    & 1.73   & 2.27   & 0.04 \\
    & {\scriptsize CIFAR-10}           & 4.19    & 1.82   & 2.73   & 0.05 \\
$\text{TC}$  & {\scriptsize F-MNIST}   & 826.67  & 149.34 & 166.65 & 3.13 \\
    & {\scriptsize CIFAR-10}           & 1304.12 & 274.61 & 244.98 & 4.59 \\ \hline \\[-1em]
\end{tabular}
\caption*{{\scriptsize
  $\text{TD}^1$: Threshold Determination in FedRecover;
  $\text{TD}^2$: Threshold Determination in Starfish;
  $\text{RS}^1$: Round Selection of \textit{Method 1};
  $\text{RS}^2$: Round Selection of \textit{Method 2};
  $\text{UE}$: Update Estimation;
  $\text{TC}$: Threshold Checking.
}}
\end{table}

\begin{table}[htb]
\small
\caption{2PC performance comparison across various datasets under different network settings on Fashion-MNIST and CIFAR-10 dataset.}
\label{tab:time_comm_FC}
\centering
\begin{tabular}{cccccc}
\hline \\[-1em]
    & Dataset & \multicolumn{1}{c}{\begin{tabular}[c]{@{}c@{}}Offline\\ Time {\tiny(s)}\end{tabular}} 
              & \multicolumn{1}{c}{\begin{tabular}[c]{@{}c@{}}Online\\ Time {\tiny(s)}\end{tabular}} 
              & \multicolumn{1}{c}{\begin{tabular}[c]{@{}c@{}}Offline\\ {\scriptsize Comm} {\tiny(GB)}\end{tabular}} 
              & \multicolumn{1}{c}{\begin{tabular}[c]{@{}c@{}}Online\\ {\scriptsize Comm} {\tiny(GB)}\end{tabular}} 
              \\ \hline \\[-1em]
LAN & {\scriptsize F-MNIST}  & 22181.12 & 4733.36  & 4141.57 & 77.73 \\
    & {\scriptsize CIFAR-10} & 33818.05 & 7767.60  & 6053.25  & 113.57 \\
WAN & {\scriptsize F-MNIST}  & 30434.06 & 4980.18  & 4177.08  & 78.39 \\
    & {\scriptsize CIFAR-10} & 46852.64 & 8408.44  & 6108.37 & 114.58 \\ \hline \\[-1em]
\end{tabular}
\end{table}



\begin{figure}[htb]
    \centering
    \begin{subfigure}{0.23\textwidth}
        \includegraphics[width=1\linewidth]{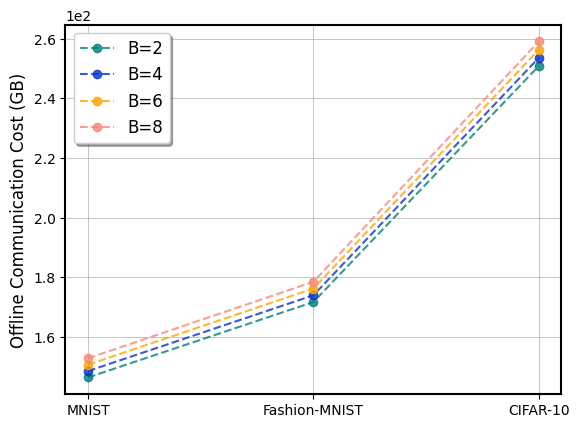}
        \caption{Offline communication}
        \label{fig:buffer_offline_comm}
    \end{subfigure}
    \begin{subfigure}{0.231\textwidth}
        \includegraphics[width=1\linewidth]{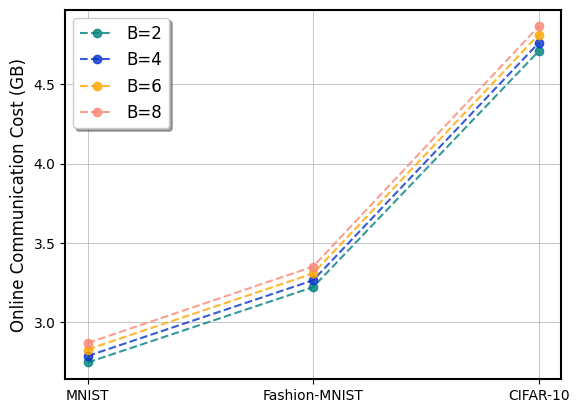}
        \caption{Online communication}
        \label{fig:buffer_online_comm}
    \end{subfigure}

    \caption{Impact of the buffer size $B$ in terms of communication cost and runtime during the offline and online phases.}
    \label{fig:buffer_size}
\end{figure}

\begin{figure}[htb]
    \centering
    \begin{subfigure}{0.23\textwidth}
        \includegraphics[width=1\linewidth]{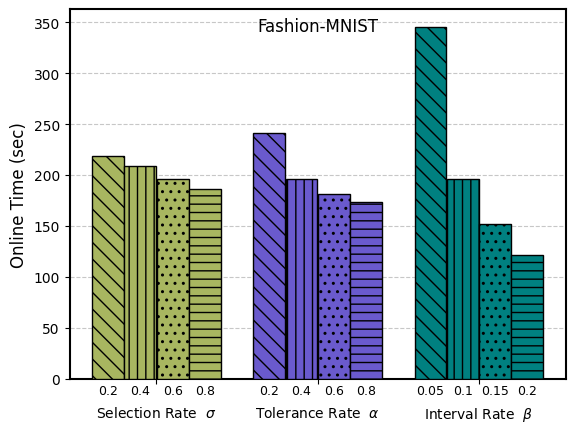}
        \caption{Online time}
        \label{fig:online_time_FashionMNIST}
    \end{subfigure}
    \begin{subfigure}{0.225\textwidth}
        \includegraphics[width=1\linewidth]{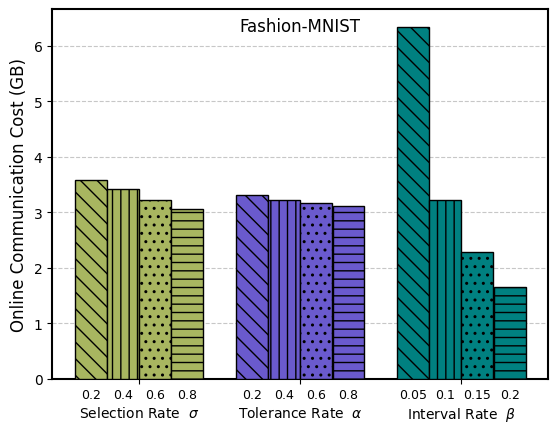}
        \caption{Online communication}
        \label{fig:online_comm_FashionMNIST}
    \end{subfigure}

    \caption{Comparison of runtime and communication cost in online phase with different selection rate $\sigma$, tolerate rate $\alpha$, and interval rate $\beta$ on Fashion-MNIST dataset.}
    \label{fig:para_time_comm_FashionMNIST}
\end{figure}

\begin{figure}[htb]
    \centering
    \begin{subfigure}{0.23\textwidth}
        \includegraphics[width=1\linewidth]{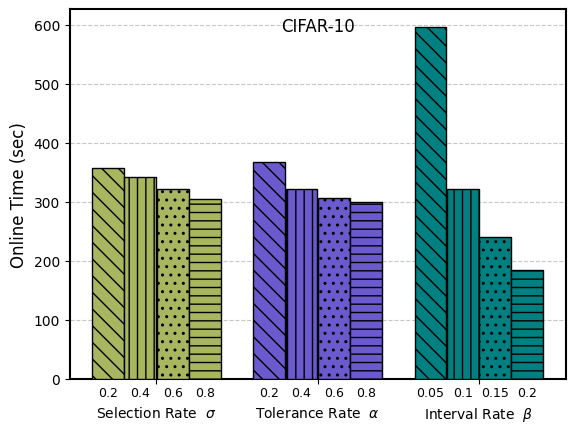}
        \caption{Online time}
        \label{fig:online_time_CIFAR10}
    \end{subfigure}
    \begin{subfigure}{0.225\textwidth}
        \includegraphics[width=1\linewidth]{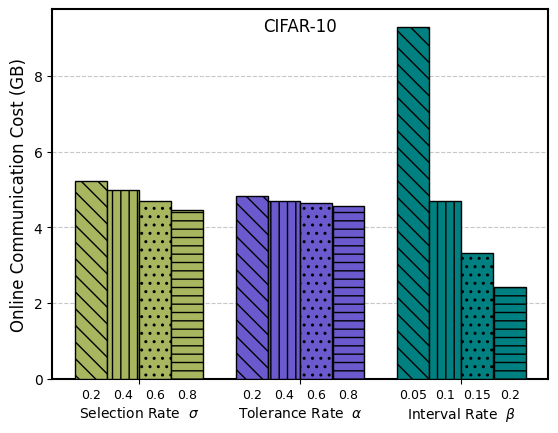}
        \caption{Online communication}
        \label{fig:online_comm_CIFAR10}
    \end{subfigure}

    \caption{Comparison of runtime and communication cost in online phase with different selection rate $\sigma$, tolerate rate $\alpha$, and interval rate $\beta$ on CIFAR-10 dataset.}
    \label{fig:para_time_comm_CIFAR10}
\end{figure}
\end{document}